\documentclass[manuscript,nolinenumber]{aastex63}
\usepackage{booktabs}

\usepackage{amsmath}
\makeatletter
\def\@to{to}
\makeatother

\shorttitle{TiO and Its Isotopologues}
\shortauthors{Witsch et al.}

\begin{document}

\title{The Rotationally Resolved Infrared Spectrum of TiO and Its Isotopologues}

\correspondingauthor{Daniel Witsch, Guido W. Fuchs}

\author{Daniel Witsch}
\email{d.witsch@physik.uni-kassel.de}
\author{Alexander A. Breier}
\author{Eileen D\"oring}
\affiliation{University of Kassel, Institute of Physics, Heinrich-Plett Str. 40, 34132 Kassel, Germany}
\author{Koichi M. T. Yamada}
\affiliation{National Metrology Institute of Japan (NMIJ), AIST, Tsukuba 305-8563, Japan}
\author{Thomas F. Giesen}
\author{Guido W. Fuchs}
\email{fuchs@physik.uni-kassel.de}
\affiliation{University of Kassel, Institute of Physics, Heinrich-Plett Str. 40, 34132 Kassel, Germany}



\begin{abstract}
In this study, we present the ro-vibrationally resolved gas-phase spectrum of the diatomic molecule TiO around 1000\,cm$^{-1}$. Molecules were produced in a laser ablation source by vaporizing a pure titanium sample in the atmosphere of gaseous nitrous oxide. Adiabatically expanded gas, containing TiO, formed a supersonic jet and was probed perpendicularly to its propagation by infrared radiation from quantum cascade lasers. Fundamental bands of $^{46-50}$TiO and vibrational hotbands of $^{48}$TiO are identified and analyzed. In a mass-independent fitting procedure combining the new infrared data with pure rotational and electronic transitions from the literature, a Dunham-like parameterization is obtained. From the present data set, the multi-isotopic analysis allows to determine the spin-rotation coupling constant $\gamma$ and the Born-Oppenheimer correction coefficient $\Delta_{\rm U_{10}}^{\mathrm{Ti}}$ for the first time. The parameter set enables to calculate the Born-Oppenheimer correction coefficients $\Delta_{\rm U_{02}}^{\mathrm{Ti}}$ and $\Delta_{\rm U_{02}}^{\mathrm{O}}$. In addition, the vibrational transition moments for the observed vibrational transitions are reported.
\end{abstract}



\section{Introduction} \label{sec:intro}
Since 1904, when the British astronomer Alfred Fowler (1868-1940) showed similarities between the spectra of Antarian stars\footnote{i.e. the spectrum of stars like Antares, $\alpha$ Herculis or $o$ Ceti} and the arc spectrum of titanium oxide (TiO), TiO has become a molecule of astrophysical relevance \citep{Fowler1904}. In the second half of the 20$^\mathrm{th}$ century, the blue-green emission in late-type stars  were assigned to electronic transitions of TiO by \cite{Merrill1962}. Since 1973, the strength of the prominent VIS-UV lines of TiO in spectra of late-type stars are used to classify stars within the Morgan-Keenan spectral classification scheme \citep{Morgan1973}. In rare cases, emission bands of TiO can be found in optical spectra of warm circumstellar environments \citep{Barnbaum1996,Kaminski2010}. Metal oxides, like TiO, are thought to play an important role in the change of the VIS-UV apparent magnitude in Mira-type variable stars at optical as well as infrared wavelengths, as has been demonstrated by \cite{Reid2002}.

Up to now, TiO signatures in stellar spectra are the topic of many observational studies. \cite{Chavez2009} detected isotopologues of TiO towards local M-dwarf stars, such as GJ699 or GJ701, at optical wavelengths. The derived isotopic abundances in these objects are similar to the natural abundance found on earth. \cite{Kaminski2013} detected pure rotational transitions from TiO towards the oxygen-rich late-type star VY Canis Majoris for the first time using the  submillimeter array (SMA) between 279.1\,GHz and 335.1\,GHz. In a follow-up study towards Mira ($o$-Ceti) at submm-wavelength, the detection of all stable titanium isotopologues was reported, with the exception of $^{47}$TiO \citep{Kaminski2017}. In these studies it was shown, that the titanium-bearing molecules, TiO and TiO$_2$, are found outside the dust forming regions. From the high abundance, the authors concluded that a substantial fraction of titanium is present as gas-phase species and not as solid dust grains. This is an indication that titanium oxides do not initiate the dust formation - as previously believed. The authors added that titanium might still support the formation of silicate dust. Recently, pure rotational transitions of vibrationally excited TiO  were identified towards the AGB stars R Dor and IK Tau by \cite{Danilovich2020}.

First evidence for signs of TiO in atmospheres of extrasolar planets (exoplanets) were found in the atmosphere of the ultra-hot jupiter WASP-121b \citep{Evans2016}, while earlier studies indicated no presence of TiO in the atmosphere of other investigated exoplanets \citep{Huitson2013,Sing2013}. TiO is proposed to contribute to thermal inversion in the atmospheres of hot Jupiters, which is in good agreement with self-consistent atmospheric models \cite{Piette2020}. Recently, an improved TiO line list was added to the ExoMol database \citep{Tennyson2016,McKemmish2019}, which contains molecules associated with atmospheres of exoplanets. Based on this line list, \cite{Pavlenko2020} investigated the spectra of the M dwarfs GJ15A and GJ15B, revealing non-solar isotopic titanium ratios deduced from synthetic spectra of the TiO isotopologues. Using the Spitzer (IR) Space Telescope, \cite{Smolders2012} published a paper, where they assigned an infrared emission band of the S-type star NP Aurigae to TiO. The spectrum was taken at a resolution of 2\,cm$^{-1}$ and no individual ro-vibrational transitions could be resolved. In the same work, further candidates of TiO emission bands were suggested in the spectra of RX Psc and V899 Aql, but could not be verified.


In terms of laboratory investigations, TiO has been intensivly studied in various experiments. An overview of the works targeting the main isotopologue $^{48}$Ti$^{16}$O is given by \cite{McKemmish2017}. In the early stages, TiO was commonly produced from modulated high-voltage arc discharge sources, like in the work of \cite{Phillips1950}. J. Phillips identified the $^1\Pi$ $-$ $^1\Delta$ and the $^1\Phi$ $-$ $^1\Delta$ bands of the main isotopologue of TiO in the spectra using a grating spectrograph. Initially, it was mistakenly assumed that the triplet electronic state $^3\Pi$ is the ground-state \citep{Lowater1929,Phillips1951}. In 1969, the $^3\Delta$ state was correctly assigned as electronic ground-state \citep{Phillips1969}. \cite{Fletcher1993} investigated the hyperfine splitting of $^{47}$TiO, that originates from the $I=5/2$ spin of the rare titanium isotope, using laser induced fluorescence spectra from the $^3\Pi$ $-$ $^3\Delta$ transition. \cite{Amiot1994,Amiot1996} performed crossed beam experiments between a molecular beam of TiO and a continuous wave tunable laser to study the $^3\Pi$ $-$ $^3\Delta$ and the $^1\Phi$ $-$ $^1\Delta$ bands at sub-Doppler re solution. The laser induced fluorescence spectra led to a comprehensive determination of the rotational molecular constants, as well as spin-orbit- and spin-spin-coupling constants, $A$ and $\lambda$. A few years later, two bands, $^3\Phi$ $-$ $^3\Delta$ and $^1\Pi$ $-$ $^1\Delta$ of the TiO main isotopologue, were identified in sunspots by \cite{Ram1996,Ram1999}. This observation was based on laboratory measurements conducted between 10,000 and 16,000\,cm$^{-1}$, where TiO was produced using a hollow cathode lamp. The same authors also improved the $^3\Delta$ ground-state rotational constants of $^{48}$TiO by including pure rotational transitions from jet cooled-measurements. \cite{Namiki2002} and \cite{Namiki2003,Namiki2004} analyzed several vibronic transitions of $^{48}$TiO of the $^3\Delta$ electronic ground-state in the optical frequency range. Measurements of the $^3\Pi$ $-$ $^3\Delta$ transition of all TiO isotopologues in a jet-cooled experiment were performed by \cite{Kobayashi2002}. Pure rotational transitions of laser ablated TiO were measured by \cite{Kania2008} in a supersonic-jet expansion utilizing millimeter-wave spectroscopy. Recently, \cite{Lincowski2016} presented a spectroscopic analysis of the rare isotopologues of TiO by means of high-resolution millimeter/submillimeter spectroscopy. From this study hyperfine constants of $^{47}$TiO and $^{49}$TiO were determined. The results agree well with those obtained by \cite{Fletcher1993}. \cite{Lincowski2016} also obtained rotational constants, including centrifugal distortion constants for spin-spin and spin-orbit parameters, for all stable titanium isotopologues. \cite{Breier2019} conducted measurements for all stable titanium isotopologues of TiO around 300\,GHz. The data were analyzed using a mass-independent Dunham approach, which allows to investigate all isotopologues in a global fit, thus combining measurements at microwave and optical wavelengths. In turn, the Dunham analysis allowed to predict frequencies for the radioactive molecule $^{44}$TiO within a sub-MHz uncertainty. This unstable isotopologue decays with a half-life of 60 years and is of particular interest in the context of young supernova remnants \citep{Siegert2015,Tsygankov2016,Austin2017}. The titanium isotope $^{44}$Ti is synthesized in core-collapse supernovae during helium burning reactions.

In this study, we present the first measurement of the infrared spectra of TiO and of all stable titanium isotopologues as well as the spectra of the vibrational hotbands of  $^{48}$TiO. In total 1034 transitions have been assigned and a line list with data of accuracy better than 10$^{-3}$\,cm$^{-1}$ was assembled to guide future astronomical observations. Transitions from vibrationally excited states of up to $v=3$ have been observed in our experiments. Together with data from the literature \citep{Breier2019,Lincowski2016,Fletcher1993,Ram1999,Amiot1994}, a mass-independent Dunham analysis, based on that presented in \cite{Breier2019}, was performed, which results in new insights into the vibrational potential of TiO.

\section{Experimental Approach}
The experiments in this work have been performed using a laser ablation technique combined with a supersonic jet expansion, intersected with infrared radiation from a quantum cascade laser (QCL), see \cite{Witsch2019} for a detailed description of the experimental setup.
Here, only a brief description is given.
A highly intense laser pulse of 7\,ns pulse duration with an output power of up to 33.5 mJ/pulse produced by a Nd:YAG laser from \emph{Continuum Lasers} (\emph{Inlite II-20 Series}) was focused onto the surface of a pure titanium sample rod and produced a hot titanium plasma ($\sim 10,000\,$K).
The sample rod with a diameter of 1\,cm and a length of 5\,cm containing titanium isotopes in natural abundances (see Table \ref{tab:transitions}) was rotated and translated in order to continuously provide a pristine surface for the ablation processes. 
A helium buffer gas containing 3\,\% of nitrous oxide (N$_2$O) as an oxygen donor was used as a carrier gas, for picking up parts of the ablated material.
A \emph{Parker Series 9 General Valve} was used to produce carrier gas pulses of  500\,$\mu$s duration.
In a subsequent reaction channel of 4\,mm length and 12\,mm $\times$ 1\,mm cross section, TiO was formed and finally adiabatically expanded into a vacuum chamber of $1.2\times10^{-1}\,$mbar background pressure. 
As a result, a supersonic jet was formed, which was probed by a quantum cascade laser (QCL) beam perpendicular to its propagation.
A Herriott-type multi-pass cell \citep{Herriott1964} guided the radiation 40 times through the jet to increase the absorption length.
The experiment was operated at a repetition rate of 20\,Hz.
The delay between the gas flow and the ablation laser pulse was controlled by a trigger pulse generator (\emph{Quantum Composers 9300 series}) to optimize the TiO yield.

In the frequency range between 984 and 992\,cm$^{-1}$, spectra were measured in a step-scan mode \citep{Witsch2019}, using a narrow linewidth distributed feedback (DFB) QCL (\emph{Alpes Lasers}).
After 40 ablation processes the QCL was stepped by $1.62\times10^{-4}$\,cm$^{-1}$. 
A liquid nitrogen cooled mercury cadmium tellurid (MCT) detector with 300\,kHz response was used to record absorption signals of TiO.

Furthermore, experiments were performed between 971\,cm$^{-1}$ and 1032\,cm$^{-1}$ using a mode-hop-free external cavity (ec) QCL from \emph{Daylight Solutions}. The infrared radiation was detected by a fast liquid nitrogen cooled 1\,GHz-response-MCT detector. By applying a 200\,kHz frequency modulation to the ec-QCL, spectral intervals of 0.03\,cm$^{-1}$ width were acquired 600 times before stepping to the next spectral interval.

All spectra were calibrated by simultaneously measuring the spectrum of an internally coupled etalon with a free spectral range of 0.006\,cm$^{-1}$ (0.01\,cm$^{-1}$ in step scan measurements) and the spectrum of methanol (CH$_3$OH) in a Herriott-type multi-pass gas cell.
The pressure in the cell was reduced until the Doppler width dominated the line broadening. 
The reference spectrum was calibrated using line positions from the Hitran database \citep{Xu2004}. 
The calibration accuracy was better than $2.8\times10^{-4}$\,cm$^{-1}$.

\section{The IR Spectrum of TiO}
\begin{figure}
	\begin{center}
		\includegraphics[scale=.5]{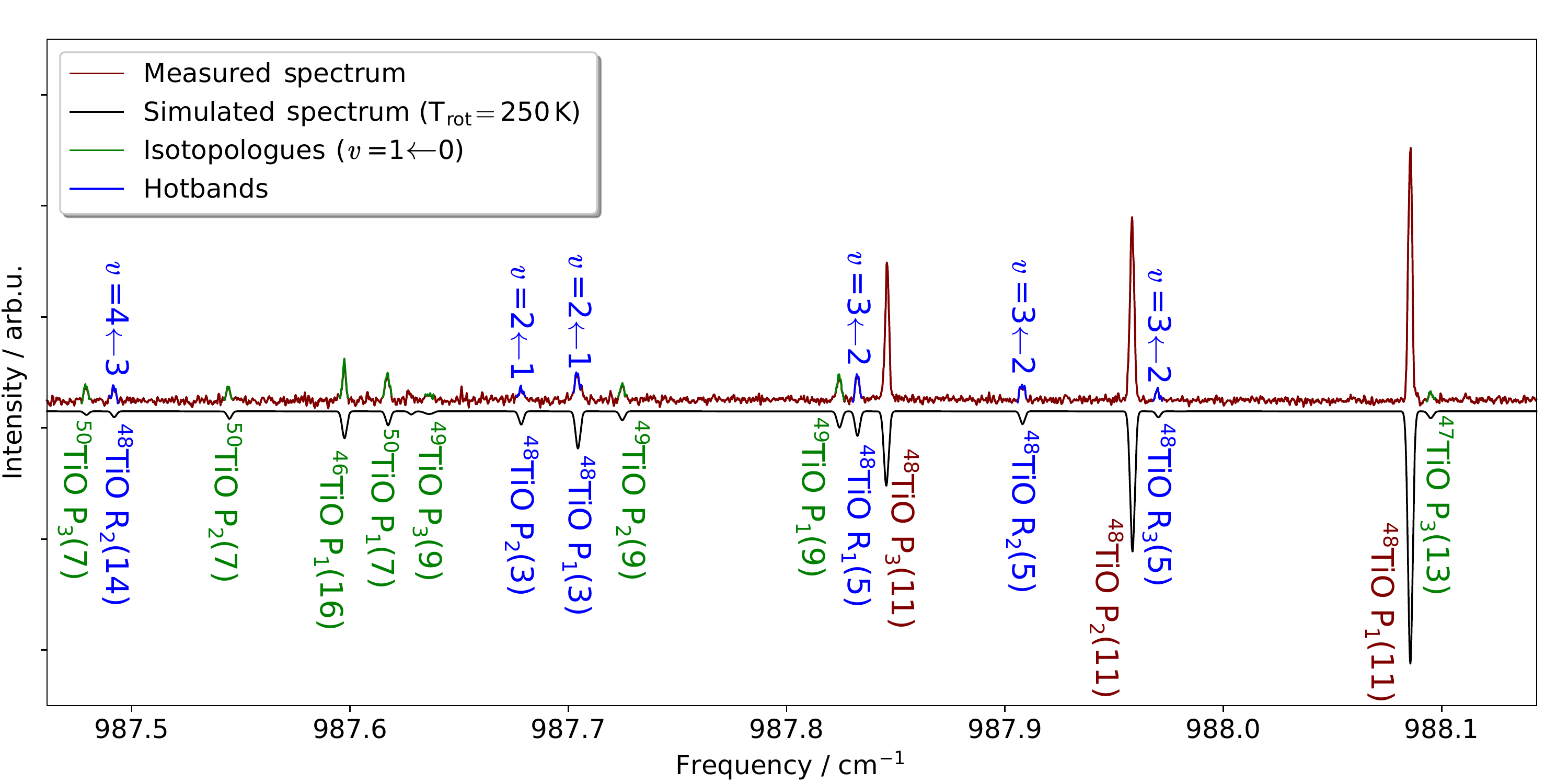}\\
		\includegraphics[scale=.5]{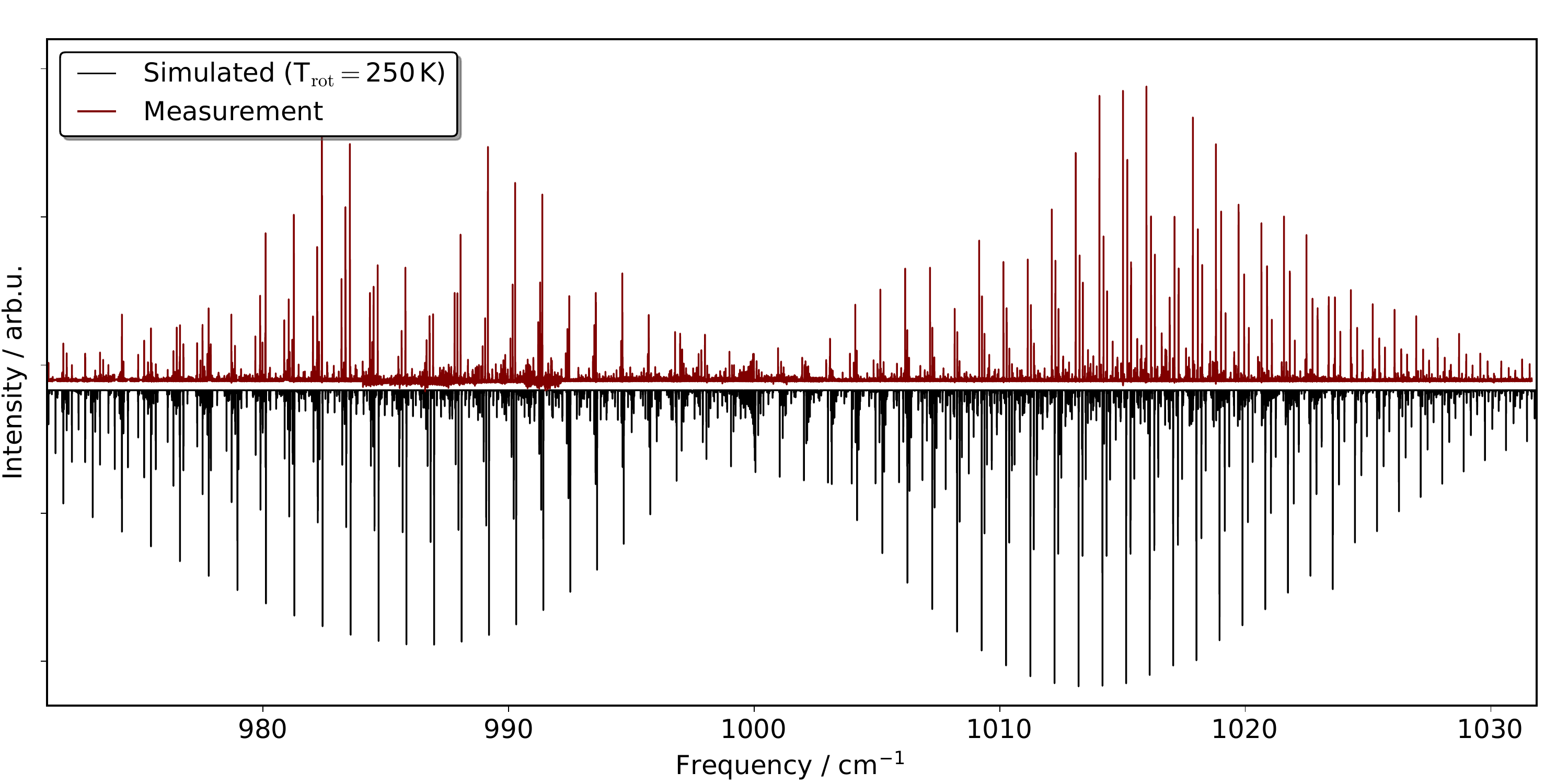}
		\caption{The infrared spectrum of TiO. Lower plot: The measured spectrum is shown (upper trace, red) together with a simulated spectrum (lower trace, black) of the fundamental bands of $^{46-50}$TiO and hotbands of $^{48}$TiO up to $v=4\leftarrow3$ assuming a rotational temperature of 247\,K. In the simulation the vibrational temperature is assumed to be 1250\,K and the isotopologues are simulated using their natural abundance values \citep{DeLaeter2003}. Upper plot: A detailed view of the spectrum around 987.8\,cm$^{-1}$. The three $\Omega$ components of the fundamental band of $^{48}$TiO are marked in red, fundamental transitions assigned to the rare isotopologues of TiO are labeled in green. Transitions assigned to the $^{48}$TiO hotbands are labeled in blue. The vibrational quantum numbers for hotband transitions are indicated.}
		\label{fig:spec}
	\end{center}
\end{figure}

In this study, we present the first rotationally resolved infrared spectrum of TiO between 971.2\,cm$^{-1}$ and 1031.9\,cm$^{-1}$, as shown in Figure \ref{fig:spec}. The laser ablation source does not provide a constant molecular yield, which causes intensity fluctuations in the observed spectrum. A total of 1034 transitions were assigned to the fundamental bands of TiO and its stable titanium isotopologues, $^{46-50}$TiO, as well as vibrational hotbands of the main isotopologue $^{48}$TiO. The typical linewidth is $2\times10^{-3}\,$cm$^{-1}$ (60\,MHz). The band centers are listed in Table \ref{tab:transitions}. A detailed list of transitions is provided in the supplemental material in Tables \ref{tab:48Ti16Ov1} to \ref{tab:50Ti16Ov1}.
 
For the main isotopologue, $^{48}$TiO, ro-vibrational transitions in the $^3\Delta$ electronic ground state of $J''$ up to 25 and 32 have been assigned to the P- and the R-branch, respectively. In addition, 63 Q-branch transitions were assigned. As listed in Table \ref{tab:transitions}, several hotband transitions of $^{48}$TiO and transitions originating from $^{46}$TiO, $^{47}$TiO, $^{49}$TiO and $^{50}$TiO are identified (see Figure \ref{fig:spec}) and appear in natural abundance. Transitions of $^{47}$TiO (nuclear spin $I=5/2$) and $^{49}$TiO ($I=7/2$) with low $J''$ values exhibit larger linewidths of up to $2.5\times10^{-3}\,$cm$^{-1}$ ($75\,$MHz), because of the unresolved hyperfine structure. Due to the spin-orbit coupling, energy levels split into three $\Omega$-components. Transitions between different $\Omega$ are prohibited by selection rules, resulting in three transitions for each $J''$ as can be seen in the upper plot in Figure \ref{fig:spec}. For $J''>6$ the splitting into three $\Omega$-components was observed for all isotopologues. However, as the total angular momentum can not be smaller than its components, only states with $J\geq\Omega$ exist, thus dissolving the triplet structure for the lowest $J''$ transitions. Furthermore, for $J''\leq 5$, some $\Omega=3$ transitions were too weak to be observed and for the lowest $J''$ values some $\Omega=2$ components could not be detected due to the unfavorable partition function.

Finally, transitions belonging to higher vibrational states ($v''\geq1$) of $^{48}$TiO were found in the observed spectrum. In total 3 hotbands were assigned ($v=2\leftarrow1$, $3\leftarrow2$ and $4\leftarrow3$) indicating that the vibrational excitation of TiO is not effectively cooled in the supersonic jet expansion -- contrary to the rotational excitation. The most populated hotband ($v=2\leftarrow1$) exhibits Q-branch transitions at $990.9$\,cm$^{-1}$, while the Q-branches from the other hotbands are too weak to be observed in our spectrum.
\begin{deluxetable}{llrrcccrr}	
	   	\caption{Number of assigned transitions in the observed spectrum of TiO, its isotopologues, and hotbands together with measured band centers $\nu$ and calculated vibrational transition moments. The natural abundance of titanium isotopes is given in percent. A detailed list of transitions is given in Tables \ref{tab:48Ti16Ov1} to \ref{tab:50Ti16Ov1}.}
\tablehead{	
\multicolumn{2}{c}{Isotopologue/} & \multicolumn{1}{c}{Natural} & & &  & \colhead{ } & \colhead{This work} & \colhead{Ref.$^b$}\\
\multicolumn{2}{c}{Transition} & \multicolumn{1}{c}{abundance$^a$} & \colhead{$\nu_{i\leftarrow i-1}$ (cm$^{-1}$)} & \colhead{P-branch} & \colhead{Q-branch} & \colhead{R-branch} & \multicolumn{2}{c}{$\left|\left<\nu_i|\mu|\nu_{i-1}\right>\right|$ (D)}
}
\startdata
$^{48}$TiO & $v=1\leftarrow0$ & 73.720(22) & 1000.0410(5) & 63 & 47 & 89 & 0.230 & 0.230\\
& $v=2\leftarrow1$ & & 990.8839(5) & 41 & 32 & 103 & 0.323 & 0.323\\
& $v=3\leftarrow2$ & & 981.7032(6) & 14 & 0 & 69 & 0.393 & 0.393\\
& $v=4\leftarrow3$ & & 972.4989(7) & 0 & 0 & 42 & 0.452 & 0.451\\
$^{46}$TiO & $v=1\leftarrow0$ & 8.249(21) & 1005.4076(5) & 73 & 0 & 69& --\;\;\; & --\;\;\;\\
$^{47}$TiO & $v=1\leftarrow0$ & 7.437(14) & 1002.6670(5) & 64 & 0 & 73& --\;\;\; & --\;\;\;\\
$^{49}$TiO & $v=1\leftarrow0$ & 5.409(10) & 997.5060(5) & 49 & 0 & 79 & --\;\;\; & --\;\;\;\\
$^{50}$TiO & $v=1\leftarrow0$ & 5.185(13) & 995.0733(5) & 48 & 0 & 79 & --\;\;\; & --\;\;\;\\
\enddata
\tablenotetext{a}{Values taken from \cite{DeLaeter2003}}
\tablenotetext{b}{Values are calculated according to \cite{McKemmish2019} using \texttt{DUO} \citep{Yurchenko2016}}
\label{tab:transitions}	
\end{deluxetable}
\section{Isotopically invariant fitting procedure}
The energy levels of the diatomic molecules, titanium monoxide and its isotopologues, can be described by the Dunham formalism \citep{Dunham1932,Dunham1932b} to obtain a mass-independent molecular parameterization, according to Equation \ref{eq:Dun}. Molecular parameters of TiO were derived from a global data set analysis using high-resolution data from \cite{Breier2019} and the here presented mid-IR measurements on multi-isotopologue ro-vibrational TiO transitions. The mass-independent molecular parameterization is described in detail by \cite{Breier2018,Breier2019}. In short, the parameter description of TiO is obtained by an adjustment procedure of the following isotopically invariant equation of the molecular parameters,
\makeatletter 
\def\@eqnnum{{\normalsize \normalcolor (\theequation)}} 
\makeatother
{
	\begin{eqnarray}
	\operatorname{X}_{v,\alpha}  &=&   \sum_{k}  \bigg\{ \eta\cdot\mu_\alpha^{\,-\frac{2l+k}{2}}\cdot\left(1+\sum_{i={\rm{Ti},\rm{O}}}\frac{m_e}{M^i_\alpha}\Delta_{\operatorname{\hat{O}}_{k,l}}^i\right)_{\text{BO}}\cdot\operatorname{\hat{O}}_{k,l}\cdot\left(v+\frac{1}{2}\right)^k  \bigg\}.
	\label{eq:Dun}
	\end{eqnarray}
} 
The isotopic invariant Dunham-like fitting parameter $\operatorname{\hat{O}}_{k,l}$ takes a central role in Equation \ref{eq:Dun}.
The index $l$ describes the expansion of molecular parameters in terms of the the angular momentum operator $\hat N^2$, $\hat N^4$, $\cdots$, $\hat N^{2l}$ for the isotopologue $\alpha$ in its vibrational state $v$. For example, the molecular parameters $B_{v,\alpha}$, $D_{v,\alpha}$ and $H_{v,\alpha}$ for $l=1,\,2\,,3$, respectively, are obtained from the Dunham parameters $\operatorname{\hat{O}}_{k,l}=U_{k,l}$. The index $k$ is the ro-vibrational coupling order ($(v+1/2)^k$). In the case of $l=1$, the Dunham parameter $U_{k,l}$ are linked to the equilibrium rotational constant ($k=0$) or the rotation-vibration interaction constants ($k=1,2,\cdots$). Analogous, Dunham-like parameters $\operatorname{\hat{O}}_{k,l}$, such as $A_{k,l}$, $\gamma_{k,l}$ and $eQq_{0_{k,l}}$, are used to describe the fine- and hyperfine-structure, e.g. the spin-orbit coupling constant $A$, the spin-rotation coupling constant $\gamma$ or the electric hyperfine-structure parameter $eQq_0$.
The coefficient $\eta$ in Equation \ref{eq:Dun} is a nucleus scaling factor  being unity for rotational and fine-structure parameters. For hyperfine parameters, $\eta$ is set to be the nuclear $g_{\text{N}}$-factor for magnetic hyperfine parameters and the electric hyperfine parameters are scaled by the electric quadrupole moment $Q$, respectively. The scaling values of TiO are given in \cite{Breier2019}.
The isotopic invariance is introduced into Equation \ref{eq:Dun} by the reduced mass $\mu_\alpha$.
To obtain a more precise description of the mass invariance, the Born-Oppenheimer breakdown (BO) correction is given as sum over parameters $\Delta_{\operatorname{\hat{O}}_{k,l}}^i$. Its mass dependency is given by the fraction of the electron mass $m_e$ and the mass of atom $M_\alpha^i$ ($i=\rm{Ti},\,\rm{O}$). In this work, the BO correction parameters $\Delta_{\operatorname{\hat{O}}_{k,l}}^i$ are determined in addition to the first-order vibrational, rotational and spin-orbital expansion terms $U_{10}$, $U_{01}$ and $A_{00}$, respectively. 

In our analysis we used the isotope masses of titanium and oxygen as published by AME2016 \citep{Wang2017}. The generalized equation is implemented in the program \texttt{PGOPHER} \citep{Western2017}. Contrary to \cite{Breier2019}, the weighting process of the various data sets is changed and the well-defined iterative re-weighting procedure introduced by \cite{Watson2003} is used to improve the reproducibility.
The \texttt{ROBUST} parameter value of \texttt{PGOPHER} was determined to be 0.1. 

One advantage of a mass-independent analysis is that the effect of the spin-rotation interaction onto the energy level is distinguishable from the centrifugal distortion of the spin-orbital effects due to their difference in mass scaling behavior \citep{Muller2015}. In this work, two different fitting sets are obtained by excluding the spin-rotation interaction (``\textit{Fit A}'') or by including this contribution (``\textit{Fit B}'') in the parameter set. Both parameter sets describe the electronic ground-state ($X^3\Delta$) of TiO with 29 mass-independent Dunham-like molecular parameters, which are shown in Table \ref{tab:parameter}. The confidence of parameter set ``\textit{Fit B}'' is emphasized and its usage for TiO is recommended. This Dunham-like parameter set corresponds to 150 effective molecular parameters which describe the vibrational states $v=0$ to $4$ of $^{48}$TiO and the vibrational states $v=0,\,1$ of the observed rare isotopologues (including the hyperfine structure of $^{47}$TiO and $^{49}$TiO). Parameters for the $A^3\Phi$ and $B^3\Pi$ state, which are obtained from the global fit, are listed in Tables \ref{tab:DunhamA} and \ref{tab:DunhamB} of the supplementary material and are not further discussed. The uncertainty of the here measured IR data is $5.2\times10^{-4}$\,cm$^{-1}$. The fitting routine within \texttt{PGOPHER} is available on request.
\startlongtable	
\begin{deluxetable}{*{1}{l}*{4}{l}*{1}{l}}
				\tablecaption{Mass-invariant molecular parameters for the $X^3\Delta$ state of TiO based on the analysis of the six stable titanium isotopologues $^{46-50}$Ti$^{16}$O and $^{48}$Ti$^{18}$O.\label{tab:parameter}}
				\tablehead{				
				Parameter & \multicolumn{2}{c}{This work}&\multicolumn{2}{c}{Ref.$^a$}&\\
				\cmidrule(lr{.75em}){2-3}\cmidrule(lr{.75em}){4-5}
				$\operatorname{\hat{O}}_{k,l}$&\multicolumn{1}{c}{Fit A}&\multicolumn{1}{c}{Fit B}&\multicolumn{1}{c}{Fit A}&\multicolumn{1}{c}{Fit B}&Units}	
				\startdata
				$U_{00}$&0.0&0.0&0.0&0.0&cm$^{-1}$\\
				$U_{10}\times 10^{-3}$&3.4949905(13)&3.4949915(13)&3.4949966(26)&3.4949959(26)& cm$^{-1}$\,u$^{1/2}$\\
				$\Delta_{\rm U_{10}}^{\rm Ti}$&0.109(31)&0.114(31)& \multicolumn{1}{l}{--} & \multicolumn{1}{l}{--} &\\
				$U_{20}\times 10^{-1}$&-5.470311(43)&-5.470315(43)&-5.47093(45)&-5.47063(46)& cm$^{-1}$\,u\\
				$U_{30}\times 10^1$&-1.6302(23)&-1.6300(23)&-1.587(22)&-1.596(22)& cm$^{-1}$\,u$^{3/2}$\\
				$U_{01}$&6.4227147(57)&6.42351367(72)&6.4227037(25)&6.4234987(74)& cm$^{-1}$\,u\\
				$\Delta_{\rm U_{01}}^{\rm Ti}$&-8.397(16)&-8.314(16)&-8.253(24)&-8.282(25)&\\
				$\Delta_{\rm U_{01}}^{\rm O}$&-6.116(25)&-9.769(31)&-6.112(8)&-9.722(29)&\\
				$U_{11}\times 10^1$&-1.255794(22)&-1.255799(21)&-1.255825(44)&-1.255631(69)& cm$^{-1}$\,u$^{3/2}$\\
				$U_{21}\times 10^3$&-1.3339(25)&-1.3337(25)&-1.3467(57)&-1.3542(61)& cm$^{-1}$\,u$^{2}$\\
				$U_{02}\times 10^5$&8.67136(21)&8.67227(26)&8.67195(38)&8.67186(38)& cm$^{-1}$\,u$^{2}$\\
				$U_{12}\times 10^6$&1.7640(47)&1.7608(47)&1.7341(67)&1.7469(81)& cm$^{-1}$\,u$^{5/2}$\\
				$U_{03}\times 10^{10}$&2.08(14)&2.29(14)&1.95(17)&2.05(17)& cm$^{-1}$\,u$^{3}$\\
				$A_{00}\times 10^{-1}$&5.046090(88)&5.0642587(62)&5.065030(11)&5.064254(14)& cm$^{-1}$\\
				$\Delta_{\rm A_{00}}^{\rm Ti}\times 10^{-2}$&3.286(15)& \multicolumn{1}{l}{--} & \multicolumn{1}{l}{--} & \multicolumn{1}{l}{--} &\\
				$A_{10}\times 10^{3}$&5.025(45)&4.245(42)&6.77(73)&4.93(84)& cm$^{-1}$\,u$^{1/2}$\\
				$A_{20}\times 10^{3}$&-10.399(53)&-10.421(54)&-10.61(51)&-10.88(53)& cm$^{-1}$\,u\\
				$A_{01}\times 10^{4}$&-3.12190(99)&1.895(24)&-1.797(10)&1.722(56)& cm$^{-1}$\,u\\
				$\Delta_{\rm A_{01}}^{\rm Ti}\times 10^{-4}$& \multicolumn{1}{l}{--} & \multicolumn{1}{l}{--} &6.429(85)& \multicolumn{1}{l}{--} &\\
				$A_{11}\times 10^{5}$&-3.958(63)& \multicolumn{1}{l}{--} &-4.25(22)&7.3(28)& cm$^{-1}$\,u$^{3/2}$\\
				$\gamma_{01}\times 10^{2}$& \multicolumn{1}{l}{--} &9.326(45)& \multicolumn{1}{l}{--} &9.00(10)& cm$^{-1}$\,u\\
				$\gamma_{11}\times 10^{2}$& \multicolumn{1}{l}{--} &0.952(12)& \multicolumn{1}{l}{--} &2.36(52)& cm$^{-1}$\,u$^{3/2}$\\
				$\gamma_{02}\times 10^{7}$& \multicolumn{1}{l}{--} &5.10(191)& \multicolumn{1}{l}{--} & \multicolumn{1}{l}{--} & cm$^{-1}$\,u$^{2}$\\
				$\lambda_{00}$&1.749446(62)&1.745255(65)&1.74974(16)&1.74584(18)& cm$^{-1}$\\
				$\lambda_{10}\times 10^{2}$&-1.5066(29)&-1.5454(22)&-1.81(11)&-2.00(12)& cm$^{-1}$\,u$^{1/2}$\\
				$\lambda_{20}\times 10^{3}$& \multicolumn{1}{l}{--} & \multicolumn{1}{l}{--} &3.19(85)&3.13(85)& cm$^{-1}$\,u\\
				$\lambda_{01}\times 10^{6}$&5.65(20)&-48.60(29)&6.64(15)&-47.76(44)& cm$^{-1}$\,u\\
				$\lambda_{11}\times 10^{8}$&4.7(10)& \multicolumn{1}{l}{--} & \multicolumn{1}{l}{--} & \multicolumn{1}{l}{--} & cm$^{-1}$\,u$^{3/2}$\\
				$a_{00}\times 10^{3}$&5.6198(21)&5.6136(22)&5.6220(45)&5.6133(46)& cm$^{-1}$\,g$_{N}^{-1}$\\
				${\Delta a}_{00}\times 10^{3}$&4.613(16)&4.647(16)&4.620(31)&4.684(31)& cm$^{-1}$\,g$_{N}^{-1}$\,u$^{1/2}$\\
				$b_{00}\times 10^{2}$&2.7470(11)&2.7658(12)&2.7481(30)&2.7653(26)& cm$^{-1}$\,g$_{N}^{-1}$\\
				$c_{00}\times 10^{3}$&-2.9640(87)&-3.1638(91)&-2.971(20)&-3.160(20)& cm$^{-1}$\,g$_{N}^{-1}$\\
				$c_{01}\times 10^{4}$&1.239(17)&1.204(17)&1.257(37)&1.205(38)& cm$^{-1}$\,g$_{N}^{-1}$\,u\\
				${eQq_0}_{00}\times 10^{3}$&-6.03184(55)&-6.03246(56)&-6.0320(11)&-6.0322(11)& cm$^{-1}$\,b$^{-1}$
				\enddata
   	\tablenotetext{a}{Values taken from \cite{Breier2019}}
\end{deluxetable}

\section{Mass-independant Dunham-like parameterization}
The addition of new mid-IR ro-vibrational transitions confirm and improve the former mass-independent Dunham-like parameterization \citep{Breier2019}, see Table \ref{tab:parameter}. This can be seen, when comparing the equilibrium  bond length of $^{48}$Ti$^{16}$O (r$_\text{e}^{48}$) with previous works, see Table \ref{tab:bond}. The here derived value of r$_\text{e}^{48}$=1.62033700(14)\,\AA\, is in perfect agreement with the single isotopologue study of \cite{Ram1999} (r$_\text{e}^{48}$=1.62033709(25)\,\AA) and with the former multi-isotopologue study of \cite{Breier2019} (r$_\text{e}^{48}$= 1.62033696(7)\,\AA). The agreement with the previous works, is also reflected by the Born-Oppenheimer corrected bond length of TiO r$_\text{e}^{\text{BO}}$, evaluated from the U$_{01}$ parameter, see Table \ref{tab:bond}. The difference between the here derived r$_\text{e}^{\text{BO}}$ and r$_\text{e}^{48}$ is solely related to BO correction terms.
   \begin{deluxetable}{lll}	
   	\caption{Comparison of bond length model values for TiO.}
   	\tablehead{	
   		\colhead{ } & \colhead{Bond length (\AA)} & \colhead{Ref.}
   	}
   	\startdata
   	r$_\text{e}^{48}$ & 1.62033709(25)  & \cite{Ram1999}\\
   	 & 1.62033696(7)  & \cite{Breier2019}\\
   	 & 1.62033700(14)  & this work\\
   	r$_\text{e}^{\text{BO}}$ & 1.61999035(93)$^\text{a}$  & \cite{Breier2019}\\
	 & 1.61998846(90)  & this work
   	\enddata
   	\tablenotetext{a}{Derived from parameter U$_{01}$ of Fit B listed in Tab\,.11 \citep{Breier2019} }
   	\label{tab:bond}
   \end{deluxetable}
\subsection{Born-Oppenheimer correction coefficients}
With the present TiO parameterization (``\textit{Fit B}''), three mass-dependent BO correction parameters, namely $\Delta^{\mathrm{Ti}}_{\mathrm{U}_{01}}$,  $\Delta^{\mathrm{O}}_{\mathrm{U}_{01}}$ and  $\Delta^{\mathrm{Ti}}_{\mathrm{U}_{10}}$ are determined, see Table \ref{tab:Dun}. By comparing the derived U$_{01}$ correction values with those derived from theoretical calculations ($\Delta^{\mathrm{Ti}}_{\mathrm{U}_{01}}=-4.9$ and $\Delta^{\mathrm{O}}_{\mathrm{U}_{01}}=-6.7$, \cite{Breier2019}), an almost constant shift of 3 to 3.4 is observed. Besides this difference in absolute values, the experimental results follow the same trend as the calculated $\Delta^{\mathrm{Ti}}_{\mathrm{U}_{01}}$ and $\Delta^{\mathrm{O}}_{\mathrm{U}_{01}}$ values, i.e., in both cases the contribution to the mass shift of rotational energy levels is stronger for oxygen than for titanium.
The same trend is also observed in the case of ZrO and HfO (Table \ref{tab:Dun}). Furthermore, the vibrational mass correction value of U$_{10}$ for titanium is derived here in the same order as expected from CO and SiO, see Table \ref{tab:Dun}.\\
\begin{deluxetable}{cccccc}	
	   	\caption{Comparison of Born-Oppenheimer correction values for diatomic molecules of the form AB, i.e., TiO, CO, SiO, ZrO and HfO.}
\tablehead{
\colhead{ } & \colhead{This work} & \colhead{Ref.$^a$} & \colhead{Ref.$^b$} & \colhead{Ref.$^c$} & \colhead{Ref.$^d$}\\
\colhead{ } & \colhead{$\mathrm{TiO}$} & \colhead{$\mathrm{ CO}$} & \colhead{$\mathrm{SiO}$} & \colhead{$\mathrm{ZrO}$} & \colhead{$\mathrm{HfO}$}
}
\startdata
$\Delta_{\mathrm{U}_{10}}^{\text{A}}$ & 0.114(31)  & 0.69547(7) & 0.567(37) & -- & --\\
$\Delta_{\mathrm{U}_{10}}^{\text{B}}$ & --  & -0.16886(7) & -- & -- & --\\
$\Delta_{\mathrm{U}_{01}}^{\text{A}}$ & -8.314(16)  & -2.0567(2) & -1.2976(44) & -4.872(39) & -3.40(57)\\
$\Delta_{\mathrm{U}_{01}}^{\text{B}}$ & -9.769(31)  & -2.1047(2) & -2.0507(16) & -6.1888(25) & -5.656(23)
\enddata
\tablenotetext{a}{Values taken from \cite{Velichko2012}}
\tablenotetext{b}{Values taken from \cite{Muller2013}}
\tablenotetext{c}{Values taken from \cite{Beaton1999}}
\tablenotetext{d}{Values taken from \cite{Lesarri2002}}
\label{tab:Dun}	
\end{deluxetable}
By using the Born-Oppenheimer mass correction values $\Delta_{\mathrm{U}_{10}}^{\text{A}}$ and $\Delta_{\mathrm{U}_{01}}^{\text{A}}$ for an atom A, \cite{Ogilvie1989} derived an empirical relation to estimate the mass correction value $\Delta_{\mathrm{U}_{02}}^{\text{A}}$ for the first order centrifugal distortion term U$_{02}$. In case of titanium it is as follows
\begin{eqnarray}
\Delta_{\rm U_{02}}^{\mathrm{Ti}}  \approx 3\Delta_{\rm U_{01}}^{\mathrm{Ti}}-2\Delta_{\rm U_{10}}^{\mathrm{Ti}},
\end{eqnarray}
yielding $\Delta_{02}^{\mathrm{Ti}}=-25.2(1)$.
By applying the Kratzer-Pekeris relation \citep{Kratzer1920,Pekeris1934} to U$_{01}$ and U$_{10}$, the mass independent first order centrifugal distortion term $U_{02}^{\text{BO}}=8.67932(3)\times 10^{-5}$\,cm$^{-1}$u$^{-2}$ is calculated according to:
\begin{eqnarray}
U_{02}^{\text{BO}}=\frac{4U_{01}^3}{U_{10}^2}.
\end{eqnarray}
In contrast to $U_{02}$, the BO correction is considered for $U_{02}^{\text{BO}}$, and consequently their combination with $\Delta_{02}^{\mathrm{Ti}}$ allows to calculate the corresponding oxygen Born-Oppenheimer correction value $\Delta_{\mathrm{U}_{02}}^{\mathrm{O}}=-15(1)$ according to Equation \ref{eq:Dun}.

\subsection{Spin-rotation coupling constant}
For a diatomic molecule in a multiplet state ($\Lambda>0$), the contribution of the spin-rotation interaction and the centrifugal correction of the spin-orbit coupling to the energy levels are only distinguishable in a multi-isotopic fitting procedure \citep{Muller2013}. In this work, the spin-rotation coupling constant $\gamma$ for the $X^3\Delta$ ground-state of TiO was determined for the first time. \cite{Brown1977} have shown that the spin-rotation interaction $\gamma\,\mathrm{\hat N}\cdot\mathrm{\hat S}$ consists of two contributions $\gamma^{(1)}$ and $\gamma^{(2)}$, with $\gamma^{(1)}$ describing the first-order dipole-dipole interaction contribution of electronic spin and the rotating charges and  $\gamma^{(2)}$ introduces the second-order spin-orbit interaction contribution (\cite{Brown1977}). The term $\gamma^{(2)}$ dominates in case of molecules with a large spin-orbit interaction. According to \cite{Lefebvre1986} and assuming a microscopic description of the spin-orbital operator in a single configuration representation for the electronic states, the spin-rotation coupling constant $\gamma$ can be estimated to be 
\begin{eqnarray}
	\gamma=\gamma^{(1)}+\gamma^{(2)}\approx2\frac{\bar{A}\cdot\bar{B}}{E_{E^3\Pi}-E_{X^3\Delta}}=\gamma^{\text{E-X}}.
	\label{eq:gamma}
\end{eqnarray}
If we consider only contributions from the two lowest lying triplet configuration state X$^3\Delta$ and E$^3\Pi$ ($E_{E^3\Pi}-E_{X^3\Delta}=11826.9548(5)$\,cm$^{-1}$) of TiO (\cite{Kobayashi2002}) and assume an average spin-orbital contribution 
of both states of $\bar{A}=94.0513(5)$\,cm$^{-1}$ as well as an average rotational constant of $\bar{B}=0.524442(4)$\,cm$^{-1}$, we obtain a spin-rotation coupling constant of $\gamma^{\text{E-X}}=250.058(2)$\,MHz from Equation \ref{eq:gamma}. This value is in fairly good agreement with our experimental value of $\gamma=236.5(11)$\,MHz suggesting that the strongest contribution of the spin-rotation coupling occurs from the interaction between the states of E$^3\Pi$ and X$^3\Delta$.

\subsection{Vibrational Transition Moments}
Finally, the vibrational transition moments for the observed vibrational transitions $\nu_{i\leftarrow i-1}$ are calculated. We have utilized the Dunham-like parameterization of TiO to generate the electronic ground-state potential of TiO using the Rydberg-Klein-Rees description \citep{Rydberg1932,Rydberg1933,Klein1932,Rees1947}. The potential constants are generated with the \texttt{RKR1} program from \cite{LeRoy2017}. The potential is combined with diagonal elements of the dipole moment curve of the $X^3\Delta$ electronic ground-state taken from \cite{McKemmish2019} to determine the vibrational transition moments with the \texttt{DUO} software \citep{Yurchenko2016}. The results are listed in Table \ref{tab:transitions} and are in excellent agreement with the values reported in the literature. Furthermore, the calculated vibrational transition moments allow to reproduce the line intensities measured in our experiments very well, as depicted in Figure \ref{fig:spec}.

\section{Conclusion}
In this work, we report on 1034 ro-vibrational transitions of TiO around 1000\,cm$^{-1}$. Accurate experimental frequency positions with  an uncertainty of better than $10^{-3}$\,cm$^{-1}$ are provided in the supplementary material, see Tables \ref{tab:48Ti16Ov1} to \ref{tab:50Ti16Ov1}. We have identified the fundamental bands of the stable isotopologues $^{46-50}$TiO, as well as the three lowest lying vibrational hotbands of the main isotopologue $^{48}$TiO. Our data can be used to guide the astronomical search for TiO at infrared frequencies. The tentative detection of TiO towards the S-type star NP Aurigae using low resolution spectra at 1000\,cm$^{-1}$ from the \emph{Spitzer Space Telescope} \citep{Smolders2012} could be reexamined, using telescopes with high spectral resolution instruments such as EXES or TEXES ($R\sim100,000$) to unambiguously verify this assignment.

The previous mass-independent Dunham-like parameterization of TiO \citep{Breier2019} is improved by including experimental mid-IR data of the present study. Our analysis yields a reliable parameterization of the spin-rotation coupling constant $\gamma$ with strong contributions occurring between the states E$^3\Pi$ and X$^3\Delta$. Furthermore, additional Born-Oppenheimer correction coefficients $\Delta_{\rm U_{01}}^{\mathrm{Ti}}$, $\Delta_{\rm U_{02}}^{\mathrm{Ti}}$ and $\Delta_{\rm U_{02}}^{\mathrm{O}}$ have been determined. This enables accurate predictions of highly excited ro-vibrational states of rare TiO isotopologues, such as $^{44}$TiO, within the uncertainty accuracy of the main isotopologue. In addition, the vibrational transition moments are calculated from the Dunham-like parameterization using a RKR potential description as input for the \texttt{DUO} software.

\section*{Acknowledgment}
This paper is dedicated to the 60$^\mathrm{th}$ birthday of Stephan Schlemmer from the Universit{\"a}t zu K{\"o}ln. His friendship and inspiring work in the field of molecular spectroscopy benefited us all. The authors gratefully acknowledge many fruitful discussions, support and advices from Stephan on various occasions. This work is supported by the Deutsche Forschungsgemeinschaft (DFG, German Research Foundation) project number 328961117 ‐ SFB 1319 ELCH and project number 326572190 - FU 715/2-1.

\section*{Supplementary Materials}
The supplementary materials contain lists of the observed infrared transitions for $^{46-50}$TiO and the molecular parameters for the electronic states $A^3\Phi$ and $B^3\Pi$.

\bibliography{./TiO}

\begin{thebibliography}{}
\expandafter\ifx\csname natexlab\endcsname\relax\def\natexlab#1{#1}\fi
\providecommand{\url}[1]{\href{#1}{#1}}
\providecommand{\dodoi}[1]{doi:~\href{http://doi.org/#1}{\nolinkurl{#1}}}
\providecommand{\doeprint}[1]{\href{http://ascl.net/#1}{\nolinkurl{http://ascl.net/#1}}}
\providecommand{\doarXiv}[1]{\href{https://arxiv.org/abs/#1}{\nolinkurl{https://arxiv.org/abs/#1}}}

\bibitem[{Amiot {et~al.}(1995)Amiot, Azaroual, Luc, \& Vetter}]{Amiot1994}
Amiot, C., Azaroual, E.~M., Luc, P., \& Vetter, R. 1995, J. Chem. Phys., 102,
  4375, \dodoi{10.1063/1.469486}

\bibitem[{Amiot {et~al.}(1996)Amiot, Cheikh, Luc, \& Vetter}]{Amiot1996}
Amiot, C., Cheikh, M., Luc, P., \& Vetter, R. 1996, J. Mol. Spectrosc., 179,
  159, \dodoi{10.1006/jmsp.1996.0194}

\bibitem[{Austin {et~al.}(2017)Austin, West, \& Heger}]{Austin2017}
Austin, S.~M., West, C., \& Heger, A. 2017, Astrophys. J. Lett., 839, L9,
  \dodoi{10.3847/2041-8213/aa68e7}

\bibitem[{Barnbaum {et~al.}(1996)Barnbaum, Omont, \& Morris}]{Barnbaum1996}
Barnbaum, C., Omont, A., \& Morris, M. 1996, Astron. Astrophys., 310, 259.
\newblock \url{http://adsabs.harvard.edu/abs/1996A%26A...310..259B}

\bibitem[{Beaton \& Gerry(1999)}]{Beaton1999}
Beaton, S.~A., \& Gerry, M. C.~L. 1999, J. Chem. Phys., 110, 10715,
  \dodoi{10.1063/1.479014}

\bibitem[{Breier {et~al.}(2018)Breier, Wa{\ss}muth, B{\"u}chling, Fuchs, Gauss,
  \& Giesen}]{Breier2018}
Breier, A.~A., Wa{\ss}muth, B., B{\"u}chling, T., {et~al.} 2018, J. Mol.
  Spectrosc., 350, 43, \dodoi{10.1016/j.jms.2018.06.001}

\bibitem[{Breier {et~al.}(2019)Breier, Wa{\ss}muth, Fuchs, Gauss, \&
  Giesen}]{Breier2019}
Breier, A.~A., Wa{\ss}muth, B., Fuchs, G.~W., Gauss, J., \& Giesen, T.~F. 2019,
  J. Mol. Spectrosc., 355, 46, \dodoi{10.1016/j.jms.2018.11.006}

\bibitem[{Brown \& Watson(1977)}]{Brown1977}
Brown, J.~M., \& Watson, J. K.~G. 1977, J. Mol. Spectrosc., 65, 65,
  \dodoi{10.1016/0022-2852(77)90358-7}

\bibitem[{Chavez \& Lambert(2009)}]{Chavez2009}
Chavez, J., \& Lambert, D.~L. 2009, Astrophys. J., 699, 1906,
  \dodoi{10.1088/0004-637X/699/2/1906}

\bibitem[{Danilovich {et~al.}(accepted 2020)Danilovich, Gottlieb, Decin,
  Richards, Lee, Kami{\'n}ski, Patel, Young, \& Menten}]{Danilovich2020}
Danilovich, T., Gottlieb, C.~A., Decin, L., {et~al.} accepted 2020, Astrophys.
  J., \dodoi{https://arxiv.org/pdf/2010.06485.pdf}

\bibitem[{{De Laeter} {et~al.}(2003){De Laeter}, B{\"o}hlke, {De Bi{\`e}vre,
  P.}, Hidaka, Peiser, Rosman, \& Taylor}]{DeLaeter2003}
{De Laeter}, J.~R., B{\"o}hlke, J.~K., {De Bi{\`e}vre, P.}, {et~al.} 2003, Pure
  Appl. Chem., 75, 683, \dodoi{10.1351/pac200375060683}

\bibitem[{Dunham(1932{\natexlab{a}})}]{Dunham1932}
Dunham, J.~L. 1932{\natexlab{a}}, Phys. Rev., 41, 713,
  \dodoi{10.1103/PhysRev.41.713}

\bibitem[{Dunham(1932{\natexlab{b}})}]{Dunham1932b}
---. 1932{\natexlab{b}}, Phys. Rev., 41, 721, \dodoi{10.1103/PhysRev.41.721}

\bibitem[{Evans {et~al.}(2016)Evans, Sing, Wakeford, Nikolov, Ballester,
  Drummond, Kataria, Gibson, Amundsen, \& Spake}]{Evans2016}
Evans, T.~M., Sing, D.~K., Wakeford, H.~R., {et~al.} 2016, Astrophys. J. Lett.,
  822, L4, \dodoi{10.3847/2041-8205/822/1/L4}

\bibitem[{Fletcher {et~al.}(1993)Fletcher, Scurlock, Jung, \&
  Steimle}]{Fletcher1993}
Fletcher, D.~A., Scurlock, C.~T., Jung, K.~Y., \& Steimle, T.~C. 1993, J. Chem.
  Phys., 99, 4288, \dodoi{10.1063/1.466082}

\bibitem[{Fowler(1904)}]{Fowler1904}
Fowler, A. 1904, Proc. R. Soc. London, 73, 219.
\newblock \url{http://www.jstor.org/stable/116773}

\bibitem[{Herriott {et~al.}(1964)Herriott, Kogelnik, \&
  Kompfner}]{Herriott1964}
Herriott, D., Kogelnik, H., \& Kompfner, R. 1964, Appl. Opt., 3, 523,
  \dodoi{10.1364/AO.3.000523}

\bibitem[{Huitson {et~al.}(2013)Huitson, Sing, Pont, Fortney, Burrows, Wilson,
  Ballester, Nikolov, Gibson, Deming, Aigrain, Evans, Henry, des Etangs,
  Showman, Vidal-Madjar, \& Zahnle}]{Huitson2013}
Huitson, C.~M., Sing, D.~K., Pont, F., {et~al.} 2013, Mon. Notices Royal
  Astron. Soc., 434, 3252, \dodoi{10.1093/mnras/stt1243}

\bibitem[{Kami{\'n}ski {et~al.}(2010)Kami{\'n}ski, Schmidt, \&
  Tylenda}]{Kaminski2010}
Kami{\'n}ski, T., Schmidt, M., \& Tylenda, R. 2010, Astron. Astrophys., 522,
  A75, \dodoi{10.1051/0004-6361/201014406}

\bibitem[{Kami{\'n}ski {et~al.}(2013)Kami{\'n}ski, Gottlieb, Menten, Patel,
  Young, Br{\"u}nken, M{\"u}ller, McCarthy, Winters, \& Decin}]{Kaminski2013}
Kami{\'n}ski, T., Gottlieb, C.~A., Menten, K.~M., {et~al.} 2013, Astron.
  Astrophys., 551, A113, \dodoi{10.1051/0004-6361/201220290}

\bibitem[{Kami{\'n}ski {et~al.}(2017)Kami{\'n}ski, M{\"u}ller, Schmidt,
  Cherchneff, Wong, Br{\"u}nken, Menten, Winters, Gottlieb, \&
  Patel}]{Kaminski2017}
Kami{\'n}ski, T., M{\"u}ller, H. S.~P., Schmidt, M.~R., {et~al.} 2017, Astron.
  Astrophys., 599, A59, \dodoi{10.1051/0004-6361/201629838}

\bibitem[{Kania {et~al.}(2008)Kania, Giesen, M{\"u}ller, Schlemmer, \&
  Br{\"u}nken}]{Kania2008}
Kania, P., Giesen, T.~F., M{\"u}ller, H. S.~P., Schlemmer, S., \& Br{\"u}nken,
  S. 2008, in {2008 33rd International Conference on Infrared, Millimeter and
  Terahertz Waves}, 1--2, \dodoi{10.1109/ICIMW.2008.4665795}

\bibitem[{Klein(1932)}]{Klein1932}
Klein, O. 1932, Zeits. f. Physik, 76, 226, \dodoi{10.1007/BF01341814}

\bibitem[{Kobayashi {et~al.}(2002)Kobayashi, Hall, Muckerman, Sears, \&
  Merer}]{Kobayashi2002}
Kobayashi, K., Hall, G.~E., Muckerman, J.~T., Sears, T.~J., \& Merer, A.~J.
  2002, J. Mol. Spectrosc., 212, 133, \dodoi{10.1006/jmsp.2002.8543}

\bibitem[{Kratzer(1920)}]{Kratzer1920}
Kratzer, A. 1920, Zeits. f. Physik, 3, 289, \dodoi{10.1007/BF01327754}

\bibitem[{{Le Roy}(2017)}]{LeRoy2017}
{Le Roy}, R.~J. 2017, J. Quant. Spectrosc. Radiat. Transfer, 186, 158,
  \dodoi{10.1016/j.jqsrt.2016.03.030}

\bibitem[{Lefebvre-Brion \& Field(1986)}]{Lefebvre1986}
Lefebvre-Brion, H., \& Field, R.~W. 1986, {Perturbations in the Spectra of
  Diatomic Molecules} (Academic, Orlando)

\bibitem[{Lesarri {et~al.}(2002)Lesarri, Suenram, \& Brugh}]{Lesarri2002}
Lesarri, A., Suenram, R.~D., \& Brugh, D. 2002, J. Chem. Phys., 117, 9651,
  \dodoi{10.1063/1.1516797}

\bibitem[{Lincowski {et~al.}(2016)Lincowski, Halfen, \& Ziurys}]{Lincowski2016}
Lincowski, A.~P., Halfen, D.~T., \& Ziurys, L.~M. 2016, Astrophys. J., 833, 9,
  \dodoi{10.3847/0004-637X/833/1/9}

\bibitem[{Lowater(1929)}]{Lowater1929}
Lowater, F. 1929, Natur, 123, 644, \dodoi{10.1038/123644b0}

\bibitem[{McKemmish {et~al.}(2019)McKemmish, Masseron, Hoeijmakers,
  P{\'e}rez-Mesa, Grimm, Yurchenko, \& Tennyson}]{McKemmish2019}
McKemmish, L.~K., Masseron, T., Hoeijmakers, H.~J., {et~al.} 2019, Mon. Notices
  Royal Astron. Soc., 488, 2836, \dodoi{10.1093/mnras/stz1818}

\bibitem[{McKemmish {et~al.}(2017)McKemmish, Masseron, Sheppard, Sandeman,
  Schofield, Furtenbacher, Cs{\'a}sz{\'a}r, Tennyson, \&
  Sousa-Silva}]{McKemmish2017}
McKemmish, L.~K., Masseron, T., Sheppard, S., {et~al.} 2017, Astrophys. J.
  Suppl. Ser., 228, 15, \dodoi{10.3847/1538-4365/228/2/15}

\bibitem[{Merrill {et~al.}(1962)Merrill, Deutsch, \& Keenan}]{Merrill1962}
Merrill, P.~W., Deutsch, A.~J., \& Keenan, P.~C. 1962, Astrophys. J., 1, 21,
  \dodoi{10.1086/147348}

\bibitem[{Morgan \& Keenan(1973)}]{Morgan1973}
Morgan, W.~W., \& Keenan, P.~C. 1973, Annu. Rev. Astron. Astrophys., 11, 29,
  \dodoi{10.1146/annurev.aa.11.090173.000333}

\bibitem[{M{\"u}ller {et~al.}(2015)M{\"u}ller, Kobayashi, Takahashi, Tomaru, \&
  Matsushima}]{Muller2015}
M{\"u}ller, H. S.~P., Kobayashi, K., Takahashi, K., Tomaru, K., \& Matsushima,
  F. 2015, J. Mol. Spectrosc., 310, 92, \dodoi{10.1016/j.jms.2014.12.002}

\bibitem[{M{\"u}ller {et~al.}(2013)M{\"u}ller, Spezzano, Bizzocchi, Gottlieb,
  Esposti, \& McCarthy}]{Muller2013}
M{\"u}ller, H. S.~P., Spezzano, S., Bizzocchi, L., {et~al.} 2013, J. Phys.
  Chem. A, 117, 13843, \dodoi{10.1021/jp408391f}

\bibitem[{Namiki \& Ito(2002)}]{Namiki2002}
Namiki, K.-i.~C., \& Ito, H. 2002, J. Mol. Spectrosc., 214, 188,
  \dodoi{10.1006/jmsp.2002.8586}

\bibitem[{Namiki {et~al.}(2003)Namiki, Ito, \& Davis}]{Namiki2003}
Namiki, K.-i.~C., Ito, H., \& Davis, S.~P. 2003, J. Mol. Spectrosc., 217, 173,
  \dodoi{10.1016/S0022-2852(02)00027-9}

\bibitem[{Namiki {et~al.}(2004)Namiki, Saitoh, \& Ito}]{Namiki2004}
Namiki, K.-i.~C., Saitoh, H., \& Ito, H. 2004, J. Mol. Spectrosc., 226, 87,
  \dodoi{10.1016/j.jms.2004.03.016}

\bibitem[{Ogilvie(1989)}]{Ogilvie1989}
Ogilvie, J.~F. 1989, Spectrosc. Lett., 22, 477,
  \dodoi{10.1080/00387018908053897}

\bibitem[{Pavlenko {et~al.}(2020)Pavlenko, Yurchenko, McKemmish, \&
  Tennyson}]{Pavlenko2020}
Pavlenko, Y.~V., Yurchenko, S.~N., McKemmish, L.~K., \& Tennyson, J. 2020,
  Astron. Astrophys., 642, A77, \dodoi{10.1051/0004-6361/202037863}

\bibitem[{Pekeris(1934)}]{Pekeris1934}
Pekeris, C.~L. 1934, Phys. Rev., 45, 98, \dodoi{10.1103/PhysRev.45.98}

\bibitem[{Phillips(1950)}]{Phillips1950}
Phillips, J.~G. 1950, Astrophys. J., 111, 314, \dodoi{10.1086/145266}

\bibitem[{Phillips(1951)}]{Phillips1951}
---. 1951, Astrophys. J., 114, 152, \dodoi{10.1086/145460}

\bibitem[{Phillips(1969)}]{Phillips1969}
---. 1969, Astrophys. J., 157, 449, \dodoi{10.1086/150079}

\bibitem[{Piette {et~al.}(2020)Piette, Madhusudhan, McKemmish, Gandhi,
  Masseron, \& Welbanks}]{Piette2020}
Piette, A. A.~A., Madhusudhan, N., McKemmish, L.~K., {et~al.} 2020, Mon.
  Notices Royal Astron. Soc., 496, 3870, \dodoi{10.1093/mnras/staa1592}

\bibitem[{Ram {et~al.}(1999)Ram, Bernath, Dulick, \& Wallace}]{Ram1999}
Ram, R.~S., Bernath, P.~F., Dulick, M., \& Wallace, L. 1999, Astrophys. J.
  Suppl. Ser., 122, 331, \dodoi{10.1086/313212}

\bibitem[{Ram {et~al.}(1996)Ram, Bernath, \& Wallace}]{Ram1996}
Ram, R.~S., Bernath, P.~F., \& Wallace, L. 1996, Astrophys. J. Suppl. Ser.,
  107, 443, \dodoi{10.1086/192370}

\bibitem[{Rees(1947)}]{Rees1947}
Rees, A. L.~G. 1947, Proc. Phys. Soc. (Lond), 59, 998,
  \dodoi{10.1088/0959-5309/59/6/310}

\bibitem[{Reid \& Goldston(2002)}]{Reid2002}
Reid, M.~J., \& Goldston, J.~E. 2002, Astrophys. J., 568, 931,
  \dodoi{10.1086/338947}

\bibitem[{Rydberg(1932)}]{Rydberg1932}
Rydberg, R. 1932, Zeits. f. Physik, 73, 376, \dodoi{10.1007/BF01341146}

\bibitem[{Rydberg(1933)}]{Rydberg1933}
---. 1933, Zeits. f. Physik, 80, 514, \dodoi{10.1007/BF02057312}

\bibitem[{Siegert {et~al.}(2015)Siegert, Diehl, Krause, \&
  Greiner}]{Siegert2015}
Siegert, T., Diehl, R., Krause, M. G.~H., \& Greiner, J. 2015, Astron.
  Astrophys., 579, A124, \dodoi{10.1051/0004-6361/201525877}

\bibitem[{Sing {et~al.}(2013)Sing, {Lecavelier des Etangs}, Fortney, Burrows,
  Pont, Wakeford, Ballester, Nikolov, Henry, Aigrain, Deming, Evans, Gibson,
  Huitson, Knutson, Showman, Vidal-Madjar, Wilson, Williamson, \&
  Zahnle}]{Sing2013}
Sing, D.~K., {Lecavelier des Etangs}, A., Fortney, J.~J., {et~al.} 2013, Mon.
  Notices Royal Astron. Soc., 436, 2956, \dodoi{10.1093/mnras/stt1782}

\bibitem[{Smolders {et~al.}(2012)Smolders, Verhoelst, Neyskens, Blommaert,
  Decin, {Van Winckel}, {Van Eck}, Sloan, Cami, Hony, {De Cat}, Menu, \&
  Vos}]{Smolders2012}
Smolders, K., Verhoelst, T., Neyskens, P., {et~al.} 2012, Astron. Astrophys.,
  543, L2, \dodoi{10.1051/0004-6361/201219520}

\bibitem[{Tennyson {et~al.}(2016)Tennyson, Yurchenko, Al-Refaie, Barton, Chubb,
  Coles, Diamantopoulou, Gorman, Hill, Lam, Lodi, McKemmish, Na, Owens,
  Polyansky, Rivlin, Sousa-Silva, Underwood, Yachmenev, \& Zak}]{Tennyson2016}
Tennyson, J., Yurchenko, S.~N., Al-Refaie, A.~F., {et~al.} 2016, J. Mol.
  Spectrosc., 327, 73, \dodoi{10.1016/j.jms.2016.05.002}

\bibitem[{Tsygankov {et~al.}(2016)Tsygankov, Krivonos, Lutovinov, Revnivtsev,
  Churazov, Sunyaev, \& Grebenev}]{Tsygankov2016}
Tsygankov, S.~S., Krivonos, R.~A., Lutovinov, A.~A., {et~al.} 2016, Mon. Not.
  R. Astron. Soc., 458, 3411, \dodoi{10.1093/mnras/stw549}

\bibitem[{Velichko {et~al.}(2012)Velichko, Mikhailenko, \&
  Tashkun}]{Velichko2012}
Velichko, T.~I., Mikhailenko, S.~N., \& Tashkun, S.~A. 2012, J. Quant.
  Spectrosc. Radiat. Transfer, 113, 1643, \dodoi{10.1016/j.jqsrt.2012.04.014}

\bibitem[{Wang {et~al.}(2017)Wang, Audi, Kondev, Huang, Naimi, \&
  Xu}]{Wang2017}
Wang, M., Audi, G., Kondev, F.~G., {et~al.} 2017, Chin. Phys. C, 41, 030003,
  \dodoi{10.1088/1674-1137/41/3/030003}

\bibitem[{Watson(2003)}]{Watson2003}
Watson, J. K.~G. 2003, J. Mol. Spectrosc., 219, 326,
  \dodoi{10.1016/S0022-2852(03)00100-0}

\bibitem[{Western(2017)}]{Western2017}
Western, C.~M. 2017, J. Quant. Spectrosc. Radiat. Transfer, 186, 221,
  \dodoi{10.1016/j.jqsrt.2016.04.010}

\bibitem[{Witsch {et~al.}(2019)Witsch, Lutter, Breier, Yamada, Fuchs, Gauss, \&
  Giesen}]{Witsch2019}
Witsch, D., Lutter, V., Breier, A.~A., {et~al.} 2019, J. Phys. Chem. A, 123,
  4168, \dodoi{10.1021/acs.jpca.9b01605}

\bibitem[{Xu {et~al.}(2004)Xu, Lees, Wang, Brown, Kleiner, \& Johns}]{Xu2004}
Xu, L.-H., Lees, R.~M., Wang, P., {et~al.} 2004, J. Mol. Spectrosc., 228, 453,
  \dodoi{10.1016/j.jms.2004.05.017}

\bibitem[{Yurchenko {et~al.}(2016)Yurchenko, Lodi, Tennysona, \&
  Stolyarovb}]{Yurchenko2016}
Yurchenko, S.~N., Lodi, L., Tennysona, J., \& Stolyarovb, A.~V. 2016, Comput.
  Phys. Commun., 202, 262, \dodoi{10.1016/j.cpc.2015.12.021}

\end{thebibliography}
\bibliographystyle{aasjournal}

\appendix
\section{Supplementary Material}

\renewcommand{\LTcapwidth}{\textwidth}
\begin{longtable}{*{7}{r}|*{7}{r}}
\caption{Observed transitions for $^{48}$TiO, $v=1\leftarrow0$. Experimental errors for line positions are better than $1\times10^{-3}$\,cm$^{-1}$.}\label{tab:48Ti16Ov1}\\
\hline
\hline
$J^\prime$ & $\Omega^\prime$ & $J^{\prime\prime}$ & $\Omega^{\prime\prime}$ & \multicolumn{1}{c}{Frequency} & \multicolumn{1}{c}{Obs-Calc} & \multicolumn{1}{c|}{$E_{\rm{lower}}$} & $J^\prime$ & $\Omega^\prime$ & $J^{\prime\prime}$ & $\Omega^{\prime\prime}$ & \multicolumn{1}{c}{Frequency} & \multicolumn{1}{c}{Obs-Calc} & \multicolumn{1}{c}{$E_{\rm{lower}}$}\\
\multicolumn{2}{c}{} & \multicolumn{2}{c}{} & \multicolumn{1}{c}{cm$^{-1}$} & \multicolumn{1}{c}{$10^{-4}$\,cm$^{-1}$} & \multicolumn{1}{c|}{K} & \multicolumn{2}{c}{} & \multicolumn{2}{c}{} & \multicolumn{1}{c}{cm$^{-1}$} & \multicolumn{1}{c}{$10^{-4}$\,cm$^{-1}$} & \multicolumn{1}{c}{K}\\
\hline
\endfirsthead
\caption{\textit{(continued)}}\\
\hline
\hline
$J^\prime$ & $\Omega^\prime$ & $J^{\prime\prime}$ & $\Omega^{\prime\prime}$ & \multicolumn{1}{c}{Frequency} & \multicolumn{1}{c}{Obs-Calc} & \multicolumn{1}{c|}{$E_{\rm{lower}}$} & $J^\prime$ & $\Omega^\prime$ & $J^{\prime\prime}$ & $\Omega^{\prime\prime}$ & \multicolumn{1}{c}{Frequency} & \multicolumn{1}{c}{Obs-Calc} & \multicolumn{1}{c}{$E_{\rm{lower}}$}\\
\multicolumn{2}{c}{} & \multicolumn{2}{c}{} & \multicolumn{1}{c}{cm$^{-1}$} & \multicolumn{1}{c}{$10^{-4}$\,cm$^{-1}$} & \multicolumn{1}{c|}{K} & \multicolumn{2}{c}{} & \multicolumn{2}{c}{} & \multicolumn{1}{c}{cm$^{-1}$} & \multicolumn{1}{c}{$10^{-4}$\,cm$^{-1}$} & \multicolumn{1}{c}{K}\\
\hline
\endhead
\hline
\multicolumn{14}{l}{\textbf{Table \thetable{}} \textit{continued on next page}}
\endfoot
\hline 
\endlastfoot
			25 & 3 & 24 & 3 & 971.264 & 1.1 & 954.0184 & 
			25 & 1 & 24 & 1 & 971.8669 & 4.76 & 748.4708\\
			24 & 2 & 23 & 2 & 972.7568 & 1.2 & 822.4359 & 
			23 & 3 & 22 & 3 & 973.7034 & 4.13 & 901.2496\\
			23 & 1 & 22 & 1 & 974.253 & 2.79 & 696.7772 & 
			22 & 3 & 21 & 3 & 974.9134 & -4.19 & 876.4754\\
			22 & 2 & 21 & 2 & 975.1595 & 1.08 & 772.2968 & 
			22 & 1 & 21 & 1 & 975.4376 & 3.41 & 672.51\\
			21 & 3 & 20 & 3 & 976.119 & 1.24 & 852.7754 & 
			21 & 2 & 20 & 2 & 976.3521 & 0.68 & 748.8236\\
			21 & 1 & 20 & 1 & 976.6167 & 6.57 & 649.2964 & 
			20 & 3 & 19 & 3 & 977.3183 & 3.03 & 830.1499\\
			20 & 2 & 19 & 2 & 977.5393 & 3.34 & 726.4151 & 
			20 & 1 & 19 & 1 & 977.7897 & 5.84 & 627.1364\\
			19 & 2 & 18 & 2 & 978.7201 & 0.33 & 705.0716 & 
			18 & 3 & 17 & 3 & 979.6993 & 4.15 & 788.1239\\
\end{longtable}

\begin{longtable}{*{7}{r}|*{7}{r}}
\caption{Observed transitions for $^{48}$TiO, $v=2\leftarrow1$. Experimental errors for line positions are better than $1\times10^{-3}$\,cm$^{-1}$.}
\label{tab:48Ti16Ov2}\\
\hline
\hline
$J^\prime$ & $\Omega^\prime$ & $J^{\prime\prime}$ & $\Omega^{\prime\prime}$ & \multicolumn{1}{c}{Frequency} & \multicolumn{1}{c}{Obs-Calc} & \multicolumn{1}{c|}{$E_{\rm{lower}}$} & $J^\prime$ & $\Omega^\prime$ & $J^{\prime\prime}$ & $\Omega^{\prime\prime}$ & \multicolumn{1}{c}{Frequency} & \multicolumn{1}{c}{Obs-Calc} & \multicolumn{1}{c}{$E_{\rm{lower}}$}\\
\multicolumn{2}{c}{} & \multicolumn{2}{c}{} & \multicolumn{1}{c}{cm$^{-1}$} & \multicolumn{1}{c}{$10^{-4}$\,cm$^{-1}$} & \multicolumn{1}{c|}{K} & \multicolumn{2}{c}{} & \multicolumn{2}{c}{} & \multicolumn{1}{c}{cm$^{-1}$} & \multicolumn{1}{c}{$10^{-4}$\,cm$^{-1}$} & \multicolumn{1}{c}{K}\\
\hline
\endfirsthead
\caption{\textit{(continued)}}\\
\hline
\hline
$J^\prime$ & $\Omega^\prime$ & $J^{\prime\prime}$ & $\Omega^{\prime\prime}$ & \multicolumn{1}{c}{Frequency} & \multicolumn{1}{c}{Obs-Calc} & \multicolumn{1}{c|}{$E_{\rm{lower}}$} & $J^\prime$ & $\Omega^\prime$ & $J^{\prime\prime}$ & $\Omega^{\prime\prime}$ & \multicolumn{1}{c}{Frequency} & \multicolumn{1}{c}{Obs-Calc} & \multicolumn{1}{c}{$E_{\rm{lower}}$}\\
\multicolumn{2}{c}{} & \multicolumn{2}{c}{} & \multicolumn{1}{c}{cm$^{-1}$} & \multicolumn{1}{c}{$10^{-4}$\,cm$^{-1}$} & \multicolumn{1}{c|}{K} & \multicolumn{2}{c}{} & \multicolumn{2}{c}{} & \multicolumn{1}{c}{cm$^{-1}$} & \multicolumn{1}{c}{$10^{-4}$\,cm$^{-1}$} & \multicolumn{1}{c}{K}\\
\hline
\endhead
\hline
\multicolumn{14}{l}{\textbf{Table \thetable{}} \textit{continued on next page}}
\endfoot
\hline 
\endlastfoot
			17 & 3 & 16 & 3 & 971.8212 & 4.61 & 1767.8228 & 
			17 & 2 & 16 & 2 & 972.0063 & 3.94 & 1664.6888\\
			17 & 1 & 16 & 1 & 972.216 & 7.99 & 1566.096 & 
			16 & 2 & 15 & 2 & 973.1647 & 6.6 & 1646.6456\\
			16 & 1 & 15 & 1 & 973.3607 & 9.54 & 1548.2539 & 
			15 & 3 & 14 & 3 & 974.1554 & 4.42 & 1732.4565\\
			15 & 2 & 14 & 2 & 974.3167 & 4.42 & 1629.6625 & 
			14 & 3 & 13 & 3 & 975.313 & -0.26 & 1716.3786\\
			14 & 2 & 13 & 2 & 975.4643 & 17.63 & 1613.7396 & 
			14 & 1 & 13 & 1 & 975.6325 & 13.79 & 1515.7159\\
			13 & 2 & 12 & 2 & 976.6038 & 8.18 & 1598.8772 & 
			13 & 1 & 12 & 1 & 976.7589 & 8.28 & 1501.0202\\
			12 & 3 & 11 & 3 & 977.612 & 6.72 & 1687.435 & 
			12 & 2 & 11 & 2 & 977.7382 & 7.96 & 1585.0755\\
			12 & 1 & 11 & 1 & 977.8802 & 10.38 & 1487.3737 & 
			11 & 3 & 10 & 3 & 978.7517 & 2.07 & 1674.5698\\
\end{longtable}

\begin{longtable}{*{7}{r}|*{7}{r}}
\caption{Observed transitions for $^{48}$TiO, $v=3\leftarrow2$. Experimental errors for line positions are better than $1\times10^{-3}$\,cm$^{-1}$.}
\label{tab:48Ti16Ov3}\\
\hline
\hline
$J^\prime$ & $\Omega^\prime$ & $J^{\prime\prime}$ & $\Omega^{\prime\prime}$ & \multicolumn{1}{c}{Frequency} & \multicolumn{1}{c}{Obs-Calc} & \multicolumn{1}{c|}{$E_{\rm{lower}}$} & $J^\prime$ & $\Omega^\prime$ & $J^{\prime\prime}$ & $\Omega^{\prime\prime}$ & \multicolumn{1}{c}{Frequency} & \multicolumn{1}{c}{Obs-Calc} & \multicolumn{1}{c}{$E_{\rm{lower}}$}\\
\multicolumn{2}{c}{} & \multicolumn{2}{c}{} & \multicolumn{1}{c}{cm$^{-1}$} & \multicolumn{1}{c}{$10^{-4}$\,cm$^{-1}$} & \multicolumn{1}{c|}{K} & \multicolumn{2}{c}{} & \multicolumn{2}{c}{} & \multicolumn{1}{c}{cm$^{-1}$} & \multicolumn{1}{c}{$10^{-4}$\,cm$^{-1}$} & \multicolumn{1}{c}{K}\\
\hline
\endfirsthead
\caption{\textit{(continued)}}\\
\hline
\hline
$J^\prime$ & $\Omega^\prime$ & $J^{\prime\prime}$ & $\Omega^{\prime\prime}$ & \multicolumn{1}{c}{Frequency} & \multicolumn{1}{c}{Obs-Calc} & \multicolumn{1}{c|}{$E_{\rm{lower}}$} & $J^\prime$ & $\Omega^\prime$ & $J^{\prime\prime}$ & $\Omega^{\prime\prime}$ & \multicolumn{1}{c}{Frequency} & \multicolumn{1}{c}{Obs-Calc} & \multicolumn{1}{c}{$E_{\rm{lower}}$}\\
\multicolumn{2}{c}{} & \multicolumn{2}{c}{} & \multicolumn{1}{c}{cm$^{-1}$} & \multicolumn{1}{c}{$10^{-4}$\,cm$^{-1}$} & \multicolumn{1}{c|}{K} & \multicolumn{2}{c}{} & \multicolumn{2}{c}{} & \multicolumn{1}{c}{cm$^{-1}$} & \multicolumn{1}{c}{$10^{-4}$\,cm$^{-1}$} & \multicolumn{1}{c}{K}\\
\hline
\endhead
\hline
\multicolumn{14}{l}{\textbf{Table \thetable{}} \textit{continued on next page}}
\endfoot
\hline 
\endlastfoot
			9 & 3 & 8 & 3 & 971.8842 & -9.69 & 2642.6613 & 
			9 & 2 & 8 & 2 & 971.9776 & -7.06 & 2540.6435\\
			8 & 1 & 7 & 1 & 973.1733 & -6.69 & 2433.9371 & 
			7 & 3 & 6 & 3 & 974.1096 & -13.7 & 2624.5381\\
			7 & 2 & 6 & 2 & 974.1823 & 5.14 & 2522.6952 & 
			7 & 1 & 6 & 1 & 974.2611 & 12.88 & 2425.5857\\
			6 & 3 & 5 & 3 & 975.2129 & -16.92 & 2617.0751 & 
			6 & 2 & 5 & 2 & 975.2747 & 2.39 & 2515.3042\\
			6 & 1 & 5 & 1 & 975.341 & 12.74 & 2418.2781 & 
			5 & 3 & 4 & 3 & 976.3116 & -4.8 & 2610.6779\\
			5 & 2 & 4 & 2 & 976.3616 & 5.66 & 2508.969 & 
			5 & 1 & 4 & 1 & 976.4151 & 14.17 & 2412.0143\\
			3 & 1 & 2 & 1 & 978.5438 & 0.37 & 2402.6183 & 
			2 & 1 & 1 & 1 & 979.5989 & -8.14 & 2399.4863\\
			1 & 1 & 2 & 1 & 983.7641 & 2.75 & 2397.3982 & 
			2 & 1 & 3 & 1 & 984.7904 & 6.83 & 2399.4863\\
\end{longtable}

\begin{longtable}{*{7}{r}|*{7}{r}}
\caption{Observed transitions for $^{48}$TiO, $v=4\leftarrow3$. Experimental errors for line positions are better than $1\times10^{-3}$\,cm$^{-1}$.}
\label{tab:48Ti16Ov4}\\
\hline
\hline
$J^\prime$ & $\Omega^\prime$ & $J^{\prime\prime}$ & $\Omega^{\prime\prime}$ & \multicolumn{1}{c}{Frequency} & \multicolumn{1}{c}{Obs-Calc} & \multicolumn{1}{c|}{$E_{\rm{lower}}$} & $J^\prime$ & $\Omega^\prime$ & $J^{\prime\prime}$ & $\Omega^{\prime\prime}$ & \multicolumn{1}{c}{Frequency} & \multicolumn{1}{c}{Obs-Calc} & \multicolumn{1}{c}{$E_{\rm{lower}}$}\\
\multicolumn{2}{c}{} & \multicolumn{2}{c}{} & \multicolumn{1}{c}{cm$^{-1}$} & \multicolumn{1}{c}{$10^{-4}$\,cm$^{-1}$} & \multicolumn{1}{c|}{K} & \multicolumn{2}{c}{} & \multicolumn{2}{c}{} & \multicolumn{1}{c}{cm$^{-1}$} & \multicolumn{1}{c}{$10^{-4}$\,cm$^{-1}$} & \multicolumn{1}{c}{K}\\
\hline
\endfirsthead
\caption{\textit{(continued)}}\\
\hline
\hline
$J^\prime$ & $\Omega^\prime$ & $J^{\prime\prime}$ & $\Omega^{\prime\prime}$ & \multicolumn{1}{c}{Frequency} & \multicolumn{1}{c}{Obs-Calc} & \multicolumn{1}{c|}{$E_{\rm{lower}}$} & $J^\prime$ & $\Omega^\prime$ & $J^{\prime\prime}$ & $\Omega^{\prime\prime}$ & \multicolumn{1}{c}{Frequency} & \multicolumn{1}{c}{Obs-Calc} & \multicolumn{1}{c}{$E_{\rm{lower}}$}\\
\multicolumn{2}{c}{} & \multicolumn{2}{c}{} & \multicolumn{1}{c}{cm$^{-1}$} & \multicolumn{1}{c}{$10^{-4}$\,cm$^{-1}$} & \multicolumn{1}{c|}{K} & \multicolumn{2}{c}{} & \multicolumn{2}{c}{} & \multicolumn{1}{c}{cm$^{-1}$} & \multicolumn{1}{c}{$10^{-4}$\,cm$^{-1}$} & \multicolumn{1}{c}{K}\\
\hline
\endhead
\hline
\multicolumn{14}{l}{\textbf{Table \thetable{}} \textit{continued on next page}}
\endfoot
\hline 
\endlastfoot
			4 & 1 & 5 & 1 & 977.5936 & 12.31 & 3388.428 & 
			4 & 2 & 5 & 2 & 977.654 & -3.24 & 3485.33\\
			5 & 2 & 6 & 2 & 978.6653 & -13.62 & 3490.5787 & 
			6 & 1 & 7 & 1 & 979.5879 & -15.3 & 3399.8455\\
			6 & 2 & 7 & 2 & 979.671 & -17.2 & 3496.8769 & 
			6 & 3 & 7 & 3 & 979.739 & -27.27 & 3598.649\\
			7 & 2 & 8 & 2 & 980.6717 & -7.27 & 3504.2247 & 
			8 & 1 & 9 & 1 & 981.5629 & 11.05 & 3415.4143\\
			9 & 2 & 10 & 2 & 982.6525 & -3.11 & 3522.0682 & 
			10 & 3 & 11 & 3 & 983.7356 & -26.87 & 3634.6814\\
			11 & 1 & 12 & 1 & 984.4751 & 7.24 & 3446.5494 & 
			11 & 2 & 12 & 2 & 984.607 & -8.45 & 3544.1081\\
			12 & 1 & 13 & 1 & 985.4325 & -3.55 & 3459.0026 & 
			12 & 2 & 13 & 2 & 985.5751 & -5.96 & 3556.7013\\
			13 & 1 & 14 & 1 & 986.3852 & 1.41 & 3472.4931 & 
			13 & 2 & 14 & 2 & 986.5368 & -4.28 & 3570.3431\\
\end{longtable}

\begin{longtable}{*{7}{r}|*{7}{r}}
\caption{Observed transitions for $^{46}$TiO, $v=1\leftarrow0$. Experimental errors for line positions are better than $1\times10^{-3}$\,cm$^{-1}$.}
\label{tab:46Ti16Ov1}\\
\hline
\hline
$J^\prime$ & $\Omega^\prime$ & $J^{\prime\prime}$ & $\Omega^{\prime\prime}$ & \multicolumn{1}{c}{Frequency} & \multicolumn{1}{c}{Obs-Calc} & \multicolumn{1}{c|}{$E_{\rm{lower}}$} & $J^\prime$ & $\Omega^\prime$ & $J^{\prime\prime}$ & $\Omega^{\prime\prime}$ & \multicolumn{1}{c}{Frequency} & \multicolumn{1}{c}{Obs-Calc} & \multicolumn{1}{c}{$E_{\rm{lower}}$}\\
\multicolumn{2}{c}{} & \multicolumn{2}{c}{} & \multicolumn{1}{c}{cm$^{-1}$} & \multicolumn{1}{c}{$10^{-4}$\,cm$^{-1}$} & \multicolumn{1}{c|}{K} & \multicolumn{2}{c}{} & \multicolumn{2}{c}{} & \multicolumn{1}{c}{cm$^{-1}$} & \multicolumn{1}{c}{$10^{-4}$\,cm$^{-1}$} & \multicolumn{1}{c}{K}\\
\hline
\endfirsthead
\caption{\textit{(continued)}}\\
\hline
\hline
$J^\prime$ & $\Omega^\prime$ & $J^{\prime\prime}$ & $\Omega^{\prime\prime}$ & \multicolumn{1}{c}{Frequency} & \multicolumn{1}{c}{Obs-Calc} & \multicolumn{1}{c|}{$E_{\rm{lower}}$} & $J^\prime$ & $\Omega^\prime$ & $J^{\prime\prime}$ & $\Omega^{\prime\prime}$ & \multicolumn{1}{c}{Frequency} & \multicolumn{1}{c}{Obs-Calc} & \multicolumn{1}{c}{$E_{\rm{lower}}$}\\
\multicolumn{2}{c}{} & \multicolumn{2}{c}{} & \multicolumn{1}{c}{cm$^{-1}$} & \multicolumn{1}{c}{$10^{-4}$\,cm$^{-1}$} & \multicolumn{1}{c|}{K} & \multicolumn{2}{c}{} & \multicolumn{2}{c}{} & \multicolumn{1}{c}{cm$^{-1}$} & \multicolumn{1}{c}{$10^{-4}$\,cm$^{-1}$} & \multicolumn{1}{c}{K}\\
\hline
\endhead
\hline
\multicolumn{14}{l}{\textbf{Table \thetable{}} \textit{continued on next page}}
\endfoot
\hline 
\endlastfoot
			29 & 2 & 28 & 2 & 971.6409 & 0.7 & 1401.2293 & 
			29 & 1 & 28 & 1 & 972.0257 & 13.08 & 1254.4901\\
			28 & 2 & 27 & 2 & 972.8882 & 2.81 & 1356.2784 & 
			28 & 1 & 27 & 1 & 973.2573 & 0.28 & 1210.0286\\
			27 & 2 & 26 & 2 & 974.1296 & 3.95 & 1312.8716 & 
			26 & 2 & 25 & 2 & 975.3644 & -3.62 & 1271.0096\\
			26 & 1 & 25 & 1 & 975.7062 & 5.34 & 1125.6935 & 
			25 & 3 & 24 & 3 & 976.3057 & 2.83 & 1381.6879\\
			25 & 2 & 24 & 2 & 976.5948 & 3.73 & 1230.6929 & 
			25 & 1 & 24 & 1 & 976.9209 & -2.72 & 1085.8208\\
			24 & 2 & 23 & 2 & 977.8185 & 2.25 & 1191.9221 & 
			24 & 1 & 23 & 1 & 978.1307 & -0.73 & 1047.4786\\
			23 & 3 & 22 & 3 & 978.7729 & 2.54 & 1304.9548 & 
			23 & 2 & 22 & 2 & 979.0367 & 3.54 & 1154.6978\\
			23 & 1 & 22 & 1 & 979.3348 & 2.2 & 1010.6672 & 
			22 & 3 & 21 & 3 & 979.9978 & 3.54 & 1268.9298\\
\end{longtable}

\begin{longtable}{*{7}{r}|*{7}{r}}
\caption{Observed transitions for $^{47}$TiO, $v=1\leftarrow0$. Experimental errors for line positions are better than $1\times10^{-3}$\,cm$^{-1}$.}
\label{tab:47Ti16Ov1}\\
\hline
\hline
$J^\prime$ & $\Omega^\prime$ & $J^{\prime\prime}$ & $\Omega^{\prime\prime}$ & \multicolumn{1}{c}{Frequency} & \multicolumn{1}{c}{Obs-Calc} & \multicolumn{1}{c|}{$E_{\rm{lower}}$} & $J^\prime$ & $\Omega^\prime$ & $J^{\prime\prime}$ & $\Omega^{\prime\prime}$ & \multicolumn{1}{c}{Frequency} & \multicolumn{1}{c}{Obs-Calc} & \multicolumn{1}{c}{$E_{\rm{lower}}$}\\
\multicolumn{2}{c}{} & \multicolumn{2}{c}{} & \multicolumn{1}{c}{cm$^{-1}$} & \multicolumn{1}{c}{$10^{-4}$\,cm$^{-1}$} & \multicolumn{1}{c|}{K} & \multicolumn{2}{c}{} & \multicolumn{2}{c}{} & \multicolumn{1}{c}{cm$^{-1}$} & \multicolumn{1}{c}{$10^{-4}$\,cm$^{-1}$} & \multicolumn{1}{c}{K}\\
\hline
\endfirsthead
\caption{\textit{(continued)}}\\
\hline
\hline
$J^\prime$ & $\Omega^\prime$ & $J^{\prime\prime}$ & $\Omega^{\prime\prime}$ & \multicolumn{1}{c}{Frequency} & \multicolumn{1}{c}{Obs-Calc} & \multicolumn{1}{c|}{$E_{\rm{lower}}$} & $J^\prime$ & $\Omega^\prime$ & $J^{\prime\prime}$ & $\Omega^{\prime\prime}$ & \multicolumn{1}{c}{Frequency} & \multicolumn{1}{c}{Obs-Calc} & \multicolumn{1}{c}{$E_{\rm{lower}}$}\\
\multicolumn{2}{c}{} & \multicolumn{2}{c}{} & \multicolumn{1}{c}{cm$^{-1}$} & \multicolumn{1}{c}{$10^{-4}$\,cm$^{-1}$} & \multicolumn{1}{c|}{K} & \multicolumn{2}{c}{} & \multicolumn{2}{c}{} & \multicolumn{1}{c}{cm$^{-1}$} & \multicolumn{1}{c}{$10^{-4}$\,cm$^{-1}$} & \multicolumn{1}{c}{K}\\
\hline
\endhead
\hline
\multicolumn{14}{l}{\textbf{Table \thetable{}} \textit{continued on next page}}
\endfoot
\hline 
\endlastfoot
			27 & 3 & 26 & 3 & 971.255 & -3.79 & 1459.3858 & 
			27 & 1 & 26 & 1 & 971.9176 & 5.07 & 1161.9564\\
			26 & 2 & 25 & 2 & 972.7939 & -3.72 & 1266.0184 & 
			26 & 1 & 25 & 1 & 973.1325 & 11.75 & 1120.7779\\
			25 & 3 & 24 & 3 & 973.7306 & -8.89 & 1376.8727 & 
			24 & 3 & 23 & 3 & 974.9599 & -8.79 & 1337.9428\\
			24 & 2 & 23 & 2 & 975.2332 & -4.04 & 1187.3642 & 
			24 & 1 & 23 & 1 & 975.5455 & 30.64 & 1042.9864\\
			23 & 3 & 22 & 3 & 976.1835 & -8.45 & 1300.5649 & 
			23 & 2 & 22 & 2 & 976.4447 & 1.26 & 1150.3438\\
			23 & 1 & 22 & 1 & 976.7397 & 3.41 & 1006.3743 & 
			22 & 3 & 21 & 3 & 977.3999 & -20.99 & 1264.7396\\
			22 & 2 & 21 & 2 & 977.6503 & 5.73 & 1114.8621 & 
			22 & 1 & 21 & 1 & 977.9311 & 6.66 & 971.285\\
			21 & 3 & 20 & 3 & 978.6135 & -3.55 & 1230.4677 & 
			21 & 2 & 20 & 2 & 978.8491 & 0.44 & 1080.9193\\
\end{longtable}

\begin{longtable}{*{7}{r}|*{7}{r}}
\caption{Observed transitions for $^{49}$TiO, $v=1\leftarrow0$. Experimental errors for line positions are better than $1\times10^{-3}$\,cm$^{-1}$.}
\label{tab:49Ti16Ov1}\\
\hline
\hline
$J^\prime$ & $\Omega^\prime$ & $J^{\prime\prime}$ & $\Omega^{\prime\prime}$ & \multicolumn{1}{c}{Frequency} & \multicolumn{1}{c}{Obs-Calc} & \multicolumn{1}{c|}{$E_{\rm{lower}}$} & $J^\prime$ & $\Omega^\prime$ & $J^{\prime\prime}$ & $\Omega^{\prime\prime}$ & \multicolumn{1}{c}{Frequency} & \multicolumn{1}{c}{Obs-Calc} & \multicolumn{1}{c}{$E_{\rm{lower}}$}\\
\multicolumn{2}{c}{} & \multicolumn{2}{c}{} & \multicolumn{1}{c}{cm$^{-1}$} & \multicolumn{1}{c}{$10^{-4}$\,cm$^{-1}$} & \multicolumn{1}{c|}{K} & \multicolumn{2}{c}{} & \multicolumn{2}{c}{} & \multicolumn{1}{c}{cm$^{-1}$} & \multicolumn{1}{c}{$10^{-4}$\,cm$^{-1}$} & \multicolumn{1}{c}{K}\\
\hline
\endfirsthead
\caption{\textit{(continued)}}\\
\hline
\hline
$J^\prime$ & $\Omega^\prime$ & $J^{\prime\prime}$ & $\Omega^{\prime\prime}$ & \multicolumn{1}{c}{Frequency} & \multicolumn{1}{c}{Obs-Calc} & \multicolumn{1}{c|}{$E_{\rm{lower}}$} & $J^\prime$ & $\Omega^\prime$ & $J^{\prime\prime}$ & $\Omega^{\prime\prime}$ & \multicolumn{1}{c}{Frequency} & \multicolumn{1}{c}{Obs-Calc} & \multicolumn{1}{c}{$E_{\rm{lower}}$}\\
\multicolumn{2}{c}{} & \multicolumn{2}{c}{} & \multicolumn{1}{c}{cm$^{-1}$} & \multicolumn{1}{c}{$10^{-4}$\,cm$^{-1}$} & \multicolumn{1}{c|}{K} & \multicolumn{2}{c}{} & \multicolumn{2}{c}{} & \multicolumn{1}{c}{cm$^{-1}$} & \multicolumn{1}{c}{$10^{-4}$\,cm$^{-1}$} & \multicolumn{1}{c}{K}\\
\hline
\endhead
\hline
\multicolumn{14}{l}{\textbf{Table \thetable{}} \textit{continued on next page}}
\endfoot
\hline 
\endlastfoot
			22 & 2 & 21 & 2 & 972.7555 & 5.92 & 769.6305 & 
			21 & 2 & 20 & 2 & 973.9413 & 1.23 & 746.277\\
			20 & 3 & 19 & 3 & 974.9027 & -10.19 & 827.7024 & 
			20 & 2 & 19 & 2 & 975.1219 & 0.83 & 723.9828\\
			20 & 1 & 19 & 1 & 975.3693 & 4.65 & 624.7359 & 
			19 & 3 & 18 & 3 & 976.0901 & -5.62 & 806.2632\\
			19 & 2 & 18 & 2 & 976.2964 & -2.26 & 702.7481 & 
			18 & 3 & 17 & 3 & 977.2709 & -8.67 & 785.8939\\
			18 & 1 & 17 & 1 & 977.6862 & 10.15 & 583.7847 & 
			17 & 3 & 16 & 3 & 978.4465 & -6.04 & 766.5946\\
			17 & 2 & 16 & 2 & 978.6287 & -1.93 & 663.4588 & 
			17 & 1 & 16 & 1 & 978.8355 & 7.77 & 564.8824\\
			16 & 3 & 15 & 3 & 979.6158 & -7.05 & 748.3659 & 
			16 & 2 & 15 & 2 & 979.7861 & -2.12 & 645.4047\\
			16 & 1 & 15 & 1 & 979.979 & 4.04 & 547.0292 & 
			15 & 3 & 14 & 3 & 980.7793 & -8.63 & 731.2079\\
\end{longtable}

\begin{longtable}{*{7}{r}|*{7}{r}}
\caption{Observed transitions for $^{50}$TiO, $v=1\leftarrow0$. Experimental errors for line positions are better than $1\times10^{-3}$\,cm$^{-1}$.}\label{tab:50Ti16Ov1}\\
\hline
\hline
$J^\prime$ & $\Omega^\prime$ & $J^{\prime\prime}$ & $\Omega^{\prime\prime}$ & \multicolumn{1}{c}{Frequency} & \multicolumn{1}{c}{Obs-Calc} & \multicolumn{1}{c|}{$E_{\rm{lower}}$} & $J^\prime$ & $\Omega^\prime$ & $J^{\prime\prime}$ & $\Omega^{\prime\prime}$ & \multicolumn{1}{c}{Frequency} & \multicolumn{1}{c}{Obs-Calc} & \multicolumn{1}{c}{$E_{\rm{lower}}$}\\
\multicolumn{2}{c}{} & \multicolumn{2}{c}{} & \multicolumn{1}{c}{cm$^{-1}$} & \multicolumn{1}{c}{$10^{-4}$\,cm$^{-1}$} & \multicolumn{1}{c|}{K} & \multicolumn{2}{c}{} & \multicolumn{2}{c}{} & \multicolumn{1}{c}{cm$^{-1}$} & \multicolumn{1}{c}{$10^{-4}$\,cm$^{-1}$} & \multicolumn{1}{c}{K}\\
\hline
\endfirsthead
\caption{\textit{(continued)}}\\
\hline
\hline
$J^\prime$ & $\Omega^\prime$ & $J^{\prime\prime}$ & $\Omega^{\prime\prime}$ & \multicolumn{1}{c}{Frequency} & \multicolumn{1}{c}{Obs-Calc} & \multicolumn{1}{c|}{$E_{\rm{lower}}$} & $J^\prime$ & $\Omega^\prime$ & $J^{\prime\prime}$ & $\Omega^{\prime\prime}$ & \multicolumn{1}{c}{Frequency} & \multicolumn{1}{c}{Obs-Calc} & \multicolumn{1}{c}{$E_{\rm{lower}}$}\\
\multicolumn{2}{c}{} & \multicolumn{2}{c}{} & \multicolumn{1}{c}{cm$^{-1}$} & \multicolumn{1}{c}{$10^{-4}$\,cm$^{-1}$} & \multicolumn{1}{c|}{K} & \multicolumn{2}{c}{} & \multicolumn{2}{c}{} & \multicolumn{1}{c}{cm$^{-1}$} & \multicolumn{1}{c}{$10^{-4}$\,cm$^{-1}$} & \multicolumn{1}{c}{K}\\
\hline
\endhead
\hline
\multicolumn{14}{l}{\textbf{Table \thetable{}} \textit{continued on next page}}
\endfoot
\hline 
\endlastfoot
			21 & 1 & 20 & 1 & 971.8864 & 5.78 & 644.3644 & 
			20 & 2 & 19 & 2 & 972.8016 & -0.51 & 721.6488\\
			19 & 3 & 18 & 3 & 973.767 & 10.05 & 804.0225 & 
			19 & 1 & 18 & 1 & 974.2022 & 3.26 & 601.5254\\
			18 & 3 & 17 & 3 & 974.943 & 20.28 & 783.7534 & 
			18 & 2 & 17 & 2 & 975.1335 & 0.48 & 680.4423\\
			18 & 1 & 17 & 1 & 975.3515 & 1.47 & 581.6713 & 
			17 & 3 & 16 & 3 & 976.1107 & 5.94 & 764.549\\
			17 & 2 & 16 & 2 & 976.2907 & -0.46 & 661.4214 & 
			17 & 1 & 16 & 1 & 976.4951 & -0.22 & 562.8612\\
			16 & 3 & 15 & 3 & 977.2739 & 3.64 & 746.4098 & 
			16 & 1 & 15 & 1 & 977.6338 & 6.65 & 545.095\\
			15 & 3 & 14 & 3 & 978.4313 & 3.06 & 729.3361 & 
			15 & 2 & 14 & 2 & 978.5881 & -0.09 & 626.5455\\
			15 & 1 & 14 & 1 & 978.7658 & 3.15 & 528.3731 & 
			14 & 3 & 13 & 3 & 979.5834 & 7.37 & 713.328\\
\end{longtable}

\begin{longtable}{llll}
\caption{Mass-invariant molecular parameters of the $A^3\Phi$ state of TiO. Two sets of parameters are obtained by excluding (``\textit{Fit A}'') and including (``\textit{Fit B}'') the spin-rotation interaction in the electronic ground state (see Table \ref{tab:parameter}). In a global fit, the infrared data obtained in this work are combined with data for $^{46-50}$Ti$^{16}$O and $^{48}$Ti$^{18}$O at optical and microwave frequencies from the literature \citep{Breier2019,Lincowski2016,Fletcher1993,Ram1999,Amiot1994}.}
\label{tab:DunhamA}\\
\hline
\hline
Parameter $\operatorname{\hat{O}}_{k,l}$ & \multicolumn{1}{c}{Fit A} & \multicolumn{1}{c}{Fit B} & Units\\
\hline
\endfirsthead
\caption{\textit{(continued)}}\\
\hline
\hline
Parameter $\operatorname{\hat{O}}_{k,l}$ & \multicolumn{1}{c}{Fit A} & \multicolumn{1}{c}{Fit B} & Units\\
\hline
\endhead
\hline
\multicolumn{4}{l}{\textbf{Table \thetable{}} \textit{continued on next page}}
\endfoot
\hline 
\endlastfoot
$U_{00}\times10^{-4}$ & 1.41635548(5) & 1.41635442(5) & cm$^{-1}$\\
$U_{10}\times10^{-3}$ & 3.004397(3) & 3.004397(3) & cm$^{-1}$u$^{1/2}$\\
$U_{20}\times10^{-1}$ & -4.5972(5) & -4.5972(5) & cm$^{-1}$u\\
$U_{30}\times10^{1}$ & -4.88(2) & -4.88(2) & cm$^{-1}$u$^{3/2}$\\
$U_{01}$ & 6.084493(4) & 6.084491(4) & cm$^{-1}$u\\
$U_{11}\times10^{1}$ & -1.3146(2) & -1.3145(2) & cm$^{-1}$u$^{3/2}$\\
$U_{21}\times10^{4}$ & -9.2(4) & -9.3(4) & cm$^{-1}$u$^{2}$\\
$U_{31}\times10^{4}$ & -1.5(2) & -1.4(2) & cm$^{-1}$u$^{5/2}$\\
\end{longtable}

\begin{longtable}{llll}
\caption{Molecular parameters of the $B^3\Pi$ state of TiO in the singly excited vibrational state $v=1$. All values are given in cm$^{-1}$. Two sets of parameters are obtained by excluding (``\textit{Fit A}'') and including (``\textit{Fit B}'') the spin-rotation interaction in the electronic ground state (see Table \ref{tab:parameter}). In a global fit, the infrared data obtained in this work are combined with data for $^{46-50}$Ti$^{16}$O and $^{48}$Ti$^{18}$O at optical and microwave frequencies from the literature \citep{Breier2019,Lincowski2016,Fletcher1993,Ram1999,Amiot1994}.}
\label{tab:DunhamB}\\
\hline
\hline
Parameter & \multicolumn{1}{c}{Fit A} & \multicolumn{1}{c}{Fit B}\\
\hline
\endfirsthead
\caption{\textit{(continued)}}\\
\hline
\hline
Parameter & \multicolumn{1}{c}{Fit A} & \multicolumn{1}{c}{Fit B}\\
\hline
\endhead
\hline
\multicolumn{3}{l}{\textbf{Table \thetable{}} \textit{continued on next page}}
\endfoot
\hline 
\endlastfoot
$T\times10^{-4}$ & 1.751591015(13) & 1.751589964(14)\\
$B\times10^{1}$ & 5.028657(2) & 5.028657(2)\\
$D\times10^{7}$ & 6.9128(8) & 6.9142(9)\\
$H\times10^{13}$ & 1.90(13) & 2.02(13)\\
$A\times10^{-1}$ & 2.07884(2) & 2.07883(2)\\
$A_{\rm D}\times10^{4}$ & -1.15(2) & -1.14(2)\\
$\gamma\times10^{2}$ & 2.491(5) & 2.490(5)\\
$\lambda\times10^{1}$ & -9.305(2) & -9.305(2)\\
$\lambda_{\rm D}\times10^{6}$ & -3.7(6) & -3.8(6)\\
$\lambda_{\rm H}\times10^{10}$ & -6.0(1.4) & -5.6(1.4)\\
$o\times10^{1}$ & -6.185(2) & -6.184(2)\\
$o_{\rm D}\times10^{6}$ & 1.6(4) & 1.3(4)\\
$o_{\rm H}\times10^{10}$ & 5.0(1.0) & 6.2(1.0)\\
$p\times10^{2}$ & 2.6123(8) & 2.6121(8)\\
$p_{\rm D}\times10^{7}$ & 1.13(3) & 1.14(3)\\
$q\times10^{4}$ & 2.963(3) & 2.964(3)\\
$q_{\rm D}\times10^{10}$ & -7.2(8) & -8.2(8)\\
\end{longtable}



\end{document}